\documentclass[conference]{IEEEtran}
\IEEEoverridecommandlockouts
\usepackage{cite}
\usepackage{amsmath,amssymb,amsfonts}
\usepackage{algorithmic}
\usepackage{graphicx}
\usepackage{textcomp}
\usepackage{xcolor}
\usepackage{pbox}
\usepackage{xspace}
\usepackage{booktabs}
\usepackage{makecell}
\usepackage{listings}
\usepackage[inline]{enumitem}
\usepackage{makecell}
\usepackage{relsize,color}
\usepackage{microtype}
\usepackage{mathtools}
\usepackage{framed}
\usepackage{multirow}
\usepackage{mdframed}
\usepackage{fancyvrb}
\usepackage{xcolor}
\usepackage{float}
\usepackage{tabularx}
\usepackage{tikz}
\usepackage{balance}
\usepackage{makecell}
 \usepackage{ltl}
 \usepackage[english]{babel}
\usepackage{blindtext}
\usepackage{etoolbox}
 \usepackage{makecell}
\usepackage[colorlinks=true,urlcolor=black,linkcolor=black,citecolor=black,bookmarks=false]{hyperref}
 
 \usepackage{longtable}

\newcommand{\numspecifications}{306}

\newcommand{\BOSCH}{BOSCH}
\newcommand{\PAL}{PAL Robotics}

\newcommand{\psalmist}{PsAlM\xspace}

\definecolor{light-gray}{gray}{0.95}

\newcommand{\totalrequirements}{245}

\usepackage[xcolor=orange]{changes}
\definechangesauthor[name={Claudio Menghi},color=orange]{CM}
\usepackage[colorinlistoftodos,prependcaption,textsize=tiny]{todonotes}

\newboolean{showcomments}
\setboolean{showcomments}{true} 
\ifthenelse{\boolean{showcomments}}
{\newcommand{\nb}[2]{
  \fcolorbox{black}{yellow}{\bfseries\sffamily\scriptsize#1}
  {\sf\small$\blacktriangleright$\textit{#2}$\blacktriangleleft$}
 }
 
}
{\newcommand{\nb}[2]{}
 
}

\newcommand{\numbermissionrequirementspectra}{$428$}

\newcommand{\missionNocompilationError}{$1216$}
\newcommand{\steponematching}{$424$}

\newcommand{\secref}[1]{Section\,\ref{#1}}
\newcommand{\figref}[1]{Fig.\,\ref{#1}}
\newcommand{\Figref}[1]{Figure\,\ref{#1}} 
\newcommand{\tabref}[1]{Table\,\ref{#1}}

\newcommand{\pattern}[9]{

\begin{figure*}
\footnotesize
\begin{mdframed}
\begin{center}
\textbf{Name:} #1
\end{center}
\textbf{Intent:} #2\\
\textbf{Template:} #3\\
\textbf{Variations:} #4\\
\textbf{Examples and Known Uses:} #5\\
\textbf{Relationships:} #6\\
\textbf{Occurrences:} #7\\
\end{mdframed}
\caption{The #1 pattern}
\label{fig:#1}
\end{figure*}
}

\definecolor{codegreen}{rgb}{0,0.6,0}
\definecolor{codegray}{rgb}{0.5,0.5,0.5}
\definecolor{codepurple}{rgb}{0.58,0,0.82}
\definecolor{backcolour}{rgb}{0.95,0.95,0.92}

\newcommand*\circled[1]{\tikz[baseline=(char.base)]{
            \node[shape=circle,draw,inner sep=2pt] (char) {#1};}}
            
            \newcolumntype{?}{!{\vrule width 1pt}}

\def\BibTeX{{\rm B\kern-.05em{\sc i\kern-.025em b}\kern-.08em
    T\kern-.1667em\lower.7ex\hbox{E}\kern-.125emX}}
\begin{document}

\title{Specification Patterns for Robotic Missions}

\author{
\IEEEauthorblockN{Claudio Menghi\textsuperscript{1},
Christos Tsigkanos\textsuperscript{2}, Patrizio Pelliccione\textsuperscript{3,4},
Carlo Ghezzi\textsuperscript{5}, Thorsten Berger\textsuperscript{4}
}
\IEEEauthorblockA{\textsuperscript{1} University of Luxembourg, Luxembourg}
\IEEEauthorblockA{\textsuperscript{2} Technische Universit{\"a}t Wien, Austria}
\IEEEauthorblockA{\textsuperscript{3} University of L’Aquila, Italy.}
\IEEEauthorblockA{\textsuperscript{4} Chalmers University of Technology | University of Gothenburg, Sweden}
\IEEEauthorblockA{\textsuperscript{5} Politecnico di Milano, Italy}
}

\maketitle
\thispagestyle{plain}
\pagestyle{plain}

\begin{abstract}
Mobile and general-purpose robots increasingly support our everyday life, requiring dependable robotics control software. Creating such software 
mainly amounts to implementing their complex behaviors known as missions.
Recognizing the need, a large number of domain-specific specification languages has been proposed. These, in addition to traditional logical languages, allow the use of formally specified missions for synthesis, verification, simulation, or guiding the implementation. For instance, the logical language LTL is commonly used by experts to specify missions, as an input for planners, which synthesize the behavior a robot should have. Unfortunately, domain-specific languages  are usually 
 tied to specific robot models, while logical languages such as LTL are difficult to use by non-experts.

We present a catalog of 22 mission specification patterns for mobile robots, together with tooling for instantiating, composing, and compiling the patterns to create mission specifications. 
The patterns provide solutions for recurrent specification problems, each of which detailing the usage intent, known uses, relationships to other patterns, and---most importantly---a template mission specification in temporal logic. 
Our tooling produces  specifications expressed in the LTL and CTL temporal logics  to be used by planners, simulators, or model checkers.
The patterns originate from 245 realistic textual mission requirements extracted from the robotics literature, and they are evaluated upon a total of 441 real-world mission requirements and 1251 mission specifications. Five of these reflect scenarios we defined with two well-known industrial partners developing human-size robots. We validated our patterns' correctness with simulators and  
 two real robots.

\end{abstract}

\section{Introduction}
\noindent
Mobile robots are 
increasingly used in complex environments
aiming at autonomously realizing missions~\cite{wrs:online}.
The rapid pace of development in robotics hardware and technology demands software that can sustain this growth~\cite{brugali2007software,Lee2008,Perez2008,Gamez2013563}.
Increasingly, robots will be used for accomplishing tasks of everyday life by end-users with no expertise and knowledge in computer science, robotics, mathematics or logics. 
Providing techniques that support robotic software development is a major software-engineering challenge~\cite{brugali2007software,4799437,Gotz:2018:RIW:3149485.3149523,luckcuck2018formal,Maoz:2018:SEC:3196558.3196561,MaozFSE}.

The mission describes the high-level tasks the robotic software must accomplish~\cite{lignos2015provably}.
Among the different ways of describing missions that were proposed in the literature~\cite{Mindstorms,Choregraph,DSLInRobotics,arkin2006missionlab,Teambots,maoz2011aspectltl,Ruscio2016}, in this work, we consider declarative specifications~\cite{Broy:1991:DSD:952786.952788}. 
These describe the final outcome the software should achieve---rather than describing how to achieve it---and are prominently used in the robotics domain~\cite{menghi2018multi,ulusoy2011optimal,fainekos2009temporal,guo2013revising,wolff2013automaton,kress2011robot,doi:10.1177/0278364914546174,DBLP:journals/corr/MaozR16,maoz2011aspectltl,maozsynthesis,MaozFSE}.
Precisely specifying missions and transforming them into a form useful for automatic processing are among the main challenges in engineering robotics software~\cite{situa,hinchey2005requirements,Kramer2007DevelopmentEF,6907489m,Maoz:2018:SEC:3196558.3196561}.
On the one hand, missions should be defined with a notation that is high-level and user-friendly~\cite{Bozhinoski2015,ding2011automatic,lignos2015provably}. 
On the other hand, to enable automatic processing, the notation should be unambiguous and provide a formal and precise description of what robots should do in terms of movements and actions~\cite{lee1997graphical,smith2001events,srinivas2013graphical}.

Typically, when engineering robotics software, the missions are first expressed using natural-language requirements. 
These are then specified using
domain-specific languages, many of which have been proposed over the last decades~\cite{Mindstorms,Choregraph,DSLInRobotics,raman2013sorry}.
These  languages are often integrated with development environments that are used to generate code that can be executed within simulators or real robots~\cite{arkin2006missionlab,Teambots,maoz2011aspectltl}. 
However, these languages are typically 
bound to specific types of robots and support a limited number and type of missions.
Other works, especially coming from the robotics domain, advocate to formally specify missions in temporal logics~\cite{maozsynthesis,doi:10.1177/0278364914546174,finucane2010ltlmop,menghi2018multi}. 
Unfortunately, defining temporal logic formulae is complicated. 
As such, 
the definition of 
mission specifications is 
laborious and error-prone, as widely recognized in the software-engineering and  robotics communities~(e.g.,\cite{6016586,endo2004usability,Maoz:2018:SEC:3196558.3196561,wei2016extended}).

Conceptually, defining a robotic mission entails two problems.
First, ambiguities in mission requirements that prevent precise and unambiguous specifications must be resolved~\cite{Lignos2015,raman2013sorry,wei2016extended}.
Consider the very simple mission requirement ``the robot shall visit the kitchen and the office.'' This can be interpreted as ``visit the kitchen'' and also that at some point the robot should ``visit the office'' or visit ``the kitchen and the office in order.'' 
This highlights the ambiguity in natural language requirements formulation, and common mistakes may be introduced when diverse interpretations are given~\cite{srinivas2013graphical,shah2015resolving,kiyavitskaya2008requirements,ringert2014requirements}.
Second, creating specifications that correctly capture requirements is hard and error prone~\cite{6016586,endo2004usability,Maoz:2018:SEC:3196558.3196561,wei2016extended}.
Assume that the correct intended behavior requires that ``the kitchen and the office are visited in order,'' which is a common mission specification problem~\cite{kress2009temporal,yoo2016online}. When transforming this requirement into a precise mission specification, an expert might come up with the following formula in temporal logic:
\begin{center}
$\phi_1=\LTLf \big((r\ in\  l_1)  \wedge  \LTLf (r\ in\ l_2 )\big)$,
\end{center}
\noindent
where $r\ in\ l_1$ and $r\ in\ l_2$ signify that robot $r$ is in  the kitchen and office, respectively, and  $\LTLf$ denotes \emph{finally}.
Now, recall that the actual requirement is that the robot reaches the kitchen \emph{before} the office.
Unfortunately, the logical formula still admits that the robot reaches the office before entering the kitchen, which may be an unintended behavior.
Mitigating this problem requires defining additional behavioral constraints.
A correct formula, among others, is the following:
\begin{center}
$\phi_2=\phi_1\wedge \big(\big( \neg (r\ in\ l_2)  \big)\ \LTLu\   (r\ in\ l_1)   \big)$,
\end{center}
\noindent
where $\LTLu$ stands for \emph{until}. The additional constraint requires the office to not be visited before the kitchen, recalling a specification pattern for temporal logics known as the \emph{absence pattern}~\cite{dwyer1999patterns}.
Rather than conceiving such specifications recurrently in an ad hoc way with the risk of introducing mistakes, engineers could 
re-use validated solutions to existing mission requirements.

Specification patterns are a popular solution to the specification problem.
While precise behavioral specifications in logical languages enable reasoning about behavioral properties~\cite{EMERSON1990995,5238617}, specification is hard and error prone~\cite{Holzmann2002,Autili2007}. The problem is exacerbated, since practitioners are often unfamiliar with the intricate syntax and semantics of logical languages~\cite{dwyer1999patterns}. For instance, Dwyer et al.~\cite{dwyer1999patterns} introduced patterns for safety properties, later extended by Grunske~\cite{grunske2008specification} and Konrad et al.~\cite{konrad2005real} to address real-time and probabilistic quality properties. Autili et al.~\cite{autili2015aligning} consolidated and organized these patterns into a comprehensive catalog. 
Bianculli et al.~\cite{bianculli2012specification} applied specification patterns to the domain of Web services.  All these patterns provide template solutions that can be used to specify the respective properties. However, none of these pattern catalogs focuses on the robotic software domain to solve the mission specification problem.

We propose a pattern catalog and supporting tooling that facilitates engineering missions for mobile robots,
which implements the original, high level idea that we had recently presented~\cite{Menghi:Idea}.
We focus on robot movement as one of the major aspects considered in the robotics domain~\cite{brooks1991intelligence,brugali2005software,brugali2007stable}, as well as on how robots perform actions as they move within their environment. For each pattern we provide usage intent, known uses, relationships to other patterns, and---most importantly---a template mission specification in temporal logic. The latter relies on LTL and CTL as the most widely used formal specification languages in robotics~\cite{menghi2018multi,ulusoy2011optimal,fainekos2009temporal,guo2013revising,wolff2013automaton,kress2011robot,doi:10.1177/0278364914546174,DBLP:journals/corr/MaozR16,maoz2011aspectltl,maozsynthesis,MaozFSE}.
The catalog has been produced by analyzing 245 natural-language mission requirements systematically retrieved from the robotics literature.
From these requirements we identified recurrent mission specification problems and conceived solutions were organized as patterns in a catalog.
Our patterns provide a formally defined vocabulary that supports robotics developers in defining mission requirements. Relying on the usage of the pattern catalog as a common vocabulary allows mitigating ambiguous natural language formulations~\cite{endo2004usability}. Our patterns also provide validated mission specifications for recurrent mission requirements, facilitating the creation of correct mission specifications~\cite{DBLP:journals/corr/MaozR16}.

We implemented the tool \psalmist  (Pattern bAsed Mission specifier)~\cite{PSALM} to further support developers in rigorous mission design.
\psalmist allows (i) specifying a mission requirement through a structured English grammar, which uses patterns as basic building blocks and operators that allow composing these patterns into complex missions, and (ii) automatically generating specifications from mission requirements. 
\psalmist\ is robot-agnostic and integrated with: Spectra~\cite{Spectra} (a robot development environment), a planner~\cite{doi:10.1177/0278364914546174}, NuSMV~\cite{cimatti1999nusmv} (a model checker), and Simbad~\cite{hugues2006simbad} (a simulator for education and research). The pattern catalog and the \psalmist tool are available in an online appendix~\cite{paperstuff}.

We evaluated the benefits obtained by the usage of our pattern support in rigorous and systematic mission design.
We collected $441$ mission requirements in natural language:  $436$ obtained from robotic development environments used by practitioners (i.e., Spectra~\cite{Spectra} and LTLMoP~\cite{finucane2010ltlmop,wei2016extended}), and five defined in collaboration with two well-known robotics companies developing commercial, human-size service robots (\BOSCH\ and \PAL).
We show that most of the mission requirements were ambiguous, expressible using the proposed patterns, and that the usage of  the patterns reduces ambiguities.
We then evaluated the coverage of mission specifications. 
We collected $1229$ LTL and $22$ CTL mission specifications, from robotic development environments used by practitioners (i.e., Spectra~\cite{Spectra} and LTLMoP~\cite{finucane2010ltlmop,wei2016extended}) and research publications (i.e.,~\cite{ruchkin2018ipl}) and show that almost  all the specifications can be obtained using the proposed patterns ($1154$ over $1251$).
We also generated the specifications for the five mission requirements defined in collaboration with the two  robotic companies and fed them into an existing planner. 
The produced plans were correctly executed by real robots, showing the benefits of the pattern support in real scenarios.

To ensure the correctness of the proposed patterns we manually inspected their template mission specifications.
We additionally tested patterns correctness on a set of $12$ randomly generated models representing buildings where the robot is deployed.
We considered  ten mission requirements (each obtained by combining three patterns), converted the mission requirements into LTL mission specifications and used those to generate robots' plans. 
We used the Simbad~\cite{hugues2006simbad} simulator, to verify that the plans satisfied the intended mission requirement. 
We subsequently generated both LTL and CTL specifications from the considered mission requirements. 
We verified that the same results are obtained when they are checked on  the randomly generated models, confirming the correspondence among the CTL and LTL specifications.

\section{Background}
\label{sec:background}
\noindent
In this section, we  present the terminology used in the remainder and introduce the temporal logics LTL and CTL we used for defining the patterns' template solutions.

Recall that for communication and further refinement, the requirements of a software system are typically expressed in natural language or informal models. Refining these requirements into more formal representations avoids ambiguity, allowing automated processing and analysis. 
Such practices also emerged in the robotics engineering domain.

\noindent
$\bullet$ \emph{Mission Requirement}: 
a description in a natural language or in a domain-specific  language of the mission (also  called ``task'') the robots must perform~\cite{4209779,Lignos2015,raman2013sorry,menghi2018multi}.\\
$\bullet$  \emph{Mission Specification}: a formulation  of the mission in a logical language with a precise semantics~
\cite{ulusoy2011optimal,fainekos2009temporal,guo2013revising,doi:10.1177/0278364914546174,DBLP:journals/corr/MaozR16,maozsynthesis
}.\\
$\bullet$  \emph{Mission Specification Problem}: 
the problem of generating a mission specification from a mission requirement.\\
$\bullet$  \emph{Mission Specification Pattern}: 
a mapping between a recurrent mission-specification problem to a template solution and a description of the usage intent, known uses, and relationships to other patterns.\\
$\bullet$  \emph{Mission Specification Pattern Catalog}:  a collection of mission specification patterns organized in a hierarchy aiding at browsing and selecting patterns, in order to support decision making during mission specification.

We consider LTL (Linear Temporal Logic)~\cite{pnueli1977temporal} and CTL (Computation Tree Logic)~\cite{ben1983temporal}, 
since they are commonly used to express mission specifications in the robotic domain and are utilized extensively by the community
(e.g.,~\cite{menghi2018multi,ulusoy2011optimal,fainekos2009temporal,guo2013revising,wolff2013automaton,kress2011robot,doi:10.1177/0278364914546174,DBLP:journals/corr/MaozR16,maoz2011aspectltl,maozsynthesis,MaozFSE}). 
A temporal logic specification can 
be used for several purposes, such as (i) for producing plans through the use of planners, (ii) for analysing the mission satisfaction though the use of model checkers, and (iii) to design a robotic application.

We now briefly recall the LTL and CTL syntax and semantics. Let $\pi$ be a set of atomic propositions, LTL's syntax is the following:

\begin{center}
(LTL)\hspace{0.1cm} $\phi ::= \tau ~|~ \neg \phi ~|~ \phi \vee \phi ~|~ \LTLx \phi ~|~ \phi\ \LTLu\ \phi$ \text{where $\tau \in \pi$}.
\label{syntaxltl}
\end{center}
The semantics of LTL is defined over an infinite sequence  of truth assignments to the propositions $\pi$. 
The formula $\LTLx \phi$ expresses that $ \phi$ is true in the
 next position in a sequence, and the formula $\phi_1\ \LTLu\ \phi_2 $ expresses the property that $\phi_1$ is true until $\phi_2$ holds.

\noindent CTL's syntax is the following:
\begin{center}
(CTL)\hspace{0.1cm} $\phi  \coloneqq \tau \mid \neg \phi \mid \phi \vee \phi \mid \exists \Phi \mid \forall \Phi $, where  
$\Phi  \coloneqq \LTLx \phi \mid \phi \LTLu \phi$ and $\tau \in \pi$.
\label{syntaxctl}
\end{center}
CTL  allows the specification of properties that predicate on a branching sequence of assignments.
Specifically,  when a position of a sequence has several successors,  CTL enables the specification of  a property that must hold for all or one of the paths that start from that position.
For this reason, CTL  includes two types of formulae: \emph{state} formulae that must hold in one position of the sequence 
 and \emph{path} formulae that predicate on paths that start from a position.
The operator $\forall$ (resp. $\exists$) asserts that $\phi$ must hold on all paths (resp. on one path)  starting from the current position, while $\LTLx$ and $\LTLu$ are defined as for LTL.

\section{Methodology}
\label{sec:methodology}
\noindent
We derived our pattern catalog in three main steps.

\textbf{Collection of Mission Requirements.}
We collected mission requirements from scientific papers in the field of robotics.
We additionally considered  the software engineering literature, but noted a general absence of robotic mission specifications.
We chose major venues based on consultation with domain experts and by considering their impact factor. 
Specifically, we analyzed mission specifications published in the four major~\cite{scholarlist}
robotics venues over the last five years, 
in line with similar studies for pattern identification~\cite{dwyer1999patterns,konrad2005real,grunske2008specification}.
We analyzed all papers published within a venue with two inclusion criteria (considered in order): (i) the paper title implies some notion of robotic movement-related concept, (ii) the paper contains at least one formulation of a mission requirement involving a robot that concerns movement. 
When the paper contained more than one mission requirement, each was considered separately.

\begin{table}[t]
 \centering
 \scriptsize
 \caption{Papers and (requirements) analyzed per venue and year}
 \label{fig:paperNumber}
\begin{tabularx}{\linewidth}{ p{3.19cm} l@{\hspace{1.5mm}} l@{\hspace{1.5mm}} l@{\hspace{1.5mm}} l@{\hspace{1.5mm}} l@{\hspace{1.5mm}} l@{\hspace{1.5mm}} l@{\hspace{1.5mm}}  }
\toprule
\textsf{Robotics Venue } &  \rotatebox{90}{\textsf{2017}} &  \rotatebox{90}{\textsf{2016}} & \rotatebox{90}{\textsf{2015}} & \rotatebox{90}{\textsf{2014}} & \rotatebox{90}{\textsf{2013}}  & \rotatebox{90}{\textsf{Total}}  \\
\midrule
Intl. Conf. Robotics\,\&\,Autom. & 9(14) &  16\,(11)	& 17\,(18)	& 27\,(22)	& 16\,(15)	& 85\,(80)  \\
Intl. J. of Robotics Research & 4(8) &  13\,(12)	& 12\,(11)	& 13\,(8)		& 17\,(12)	& 59 \,(51) \\
Trans. on Robotics	 & 2(6) &  12\,(9)	& 5\,(1)		& 8\,(2)	& 4\,(2)		& 31 \,(20) \\
Intl. Conf. on Int. Robots\,\&\,Sys. & 10(23)	& 55\,(26)	& 13\,(8)		& 20\,(16)	& 33\,(21)	& 131\,(94)  \\
\bottomrule
\end{tabularx}
\vspace{-0.5cm}
\end{table}

 \begin{figure*}
 \scriptsize
\begin{tikzpicture}[sibling distance=7em,
  every node/.style = {shape=rectangle, rounded corners,
    draw, align=center,
    top color=white, bottom color=white},
      level 1/.style={sibling distance=14em, level distance=0.8cm},
  level 2/.style={sibling distance=15em},
  level 3/.style={sibling distance=4.8em},
  level 4/.style={sibling distance=10em}]]
  \node (A) []{\small Robotic Missions Specification Patterns};
   \node (B) [right=0.7cm of A] {\small Avoidance/\\ \small Invariant};
   \node (C) [above left=0.2cm and 1cm of B]{\small  Conditional/Limited};
	\node[above right=0.4cm and 0.9cm of C,top color=lightgray, bottom color=lightgray, anchor=north] (D) {\small  Past \\ \small  avoidance};
    \node[above=0.8cm of C,top color=lightgray, bottom color=lightgray, anchor=north] (E) {\small  Global\\ \small  avoidance};
     \node[above left=0.4cm and 0.9cm of C,top color=lightgray, bottom color=lightgray, anchor=north] (F) {\small  Future\\ \small  avoidance};
     \node (G)  [right=0.9cm of B] {\small  Restricted};
		\node[below=0.2cm  of G, top color=lightgray, bottom color=lightgray, anchor=north] (H) {\small Lower\\ \small Restricted\\ \small Avoidance} ;
    \node[above=1.1cm of G,top color=lightgray, bottom color=lightgray, anchor=north] (I) {\small Exact\\ \small Restricted\\ \small Avoidance} ;
    \node[above right=0.3cm and 1cm of G,top color=lightgray, bottom color=lightgray, anchor=north] (L) {\small Upper\\ \small Restricted\\ \small Avoidance }; 

     \node (N) [ left=1.7cm of A] {\small Trigger};
      \node[above right=0.1cm and 0.5cm of N,top color=lightgray, bottom color=lightgray] (O) {\small Wait };
       \node (P)  [above=0.2cm of N]{\small Reaction };
        \node[above right=1cm and 0.9cm of P,top color=lightgray, bottom color=lightgray, anchor=north] (Q) {\small Instant.\\ \small Reaction };
         \node[above=1cm of P,top color=lightgray, bottom color=lightgray, anchor=north] (R) {\small Delayed\\ \small Reaction
         }; 
          \node[above left=1cm and 0.9cm of P,top color=lightgray, bottom color=lightgray, anchor=north,dashed] (FR) {\small Fast\\ \small Reaction };  
         \node (REACT)  [left=0.7cm of N,dashed]{\small Bind };     
         \node[below left=0.2cm and 0.5cm of REACT,top color=lightgray, bottom color=lightgray, anchor=north,dashed] (BR) {\small Bound\\ \small Reaction };  
         \node[below right=0.2cm and 0.5cm of REACT,top color=lightgray, bottom color=lightgray, anchor=north,dashed] (BD) {\small Bound\\ \small Delay };  
	\node (S)[below=0.2cm of A] {\small Core Movement Patterns};
    \node[below left=0.1cm and 2cm of S] (T){\small Coverage};
    \node[below left=0.5cm and 2.5cm of T,top color=lightgray, bottom color=lightgray] (U) {\small Visit};
 	\node[below left=0.5cm and 0.5cm of T,top color=lightgray, bottom color=lightgray] (V) {\small Sequenced \\ \small Visit};    
     \node[below right=0.5cm and -1.2cm of T,top color=lightgray, bottom color=lightgray] (Z) {\small Ordered \\ \small Visit};      
      \node[below right=0.5cm and 0.5cm of T,top color=lightgray, bottom color=lightgray] (ZA) {\small Strict\\ \small Ordered\\ \small Visit}; 
     \node[below right=0.5cm and 2.5cm of T,top color=lightgray, bottom color=lightgray] (ZB) {\small Fair\\ \small Visit};
      \node[below right=0.1cm and 2cm of S] (ZC) {\small Surveillance};
      \node[below left=0.5cm and 2cm of ZC,top color=lightgray, bottom color=lightgray] (ZD) {\small Patrolling};
      \node[below left=0.5cm and 0cm of ZC,top color=lightgray, bottom color=lightgray] (ZE) {\small Sequenced\\ \small Patrolling};
       \node[below right=0.5cm and -1.2cm of ZC,top color=lightgray, bottom color=lightgray] (ZF) {\small Ordered\\ \small Patrolling};
        \node[below right=0.5cm and 0.5cm of ZC,top color=lightgray, bottom color=lightgray] (ZG) {\small Strict\\ \small  Ordered \\ \small Patrolling};
       \node[below right=0.5cm and 2.5cm of ZC,top color=lightgray, bottom color=lightgray,sibling distance=4em] (ZH) {\small Fair\\ \small Patrolling};
         
         \path (T) edge   (U)
       						edge   (V)
       						edge (Z)
       						edge (ZA)
       						edge (ZB);
         \path (ZC) edge   (ZD)
       						edge   (ZE)
       						edge (ZF)
       						edge (ZG)
       						edge (ZH);
       \path (A) edge   (B)
       						edge   (N)
       						edge (S);
       \path (S) edge   (T)
       						edge   (ZC);
       \path (B) 	edge   (C)
       					edge    (G);
       \path (C) 	edge   (D)
       					edge    (E)
       					edge    (F);
        \path (G) 	edge   (H)
       					edge    (I)
       					edge    (L);
         \path (N) 	edge   (O)
       					edge    (P)
       					edge (REACT);
       	\path (REACT) edge (BR)
       							edge (BD);
       	   \path (P) 	edge   (Q)
       					edge    (R)
       					edge    (FR);

\end{tikzpicture}
\caption{Mission specification pattern catalog. Filled nodes: patterns, non-filled nodes: categories.
}
\label{fig:specificationPatternSystem}
\end{figure*}
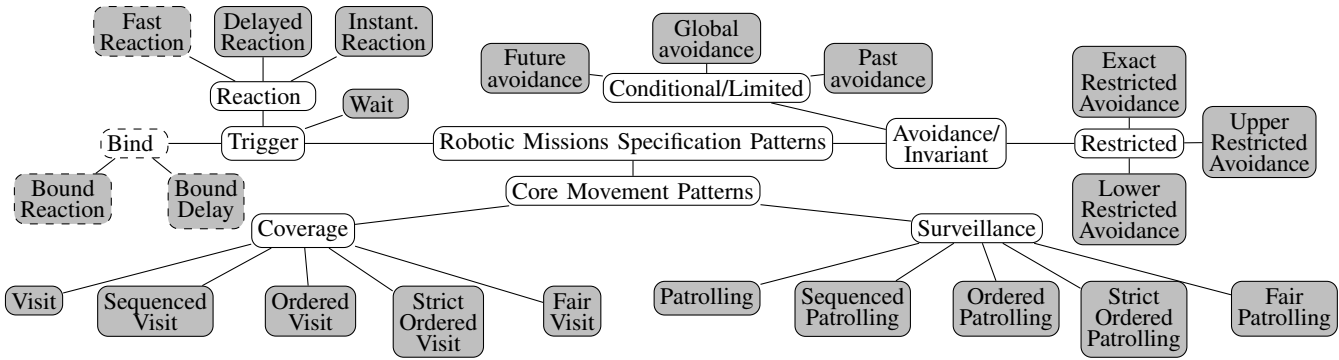

Altogether we obtained \numspecifications\ papers, 
through which, matching our inclusion criteria, we obtained \totalrequirements\ mission requirements.
\tabref{fig:paperNumber} shows the venues included in our analysis, together with the number of scientific publications and mission requirements obtained per year.
The considered software engineering venues (ICSE, FSE, and ASE) are not present, since they did not contain any paper matching the inclusion criteria.

\textbf{Identification of Mission Specification Problems.} We identified these problems as follows.

{\scshape{(Step.1)}} We divided the collected mission requirements  among two of the authors, who labeled them  with  keywords that describe the  mission specification problems they describe.
For example, the mission requirement ``The robot has to autonomously patrol the site and measure
the state of valve levers and dial gauges at four checkpoints in order to decide if some machines need to be shut down'' (occuring in Schillinger et al.~\cite{schillinger2016human}) was associated with the keywords ``patrol,'' since the robot has to patrol the site, and ``instantaneous reaction,'' since when a valve is reached its level must be checked.

{\scshape{(Step.2)}} We  created a graph structure representing semantic relations between keywords.
Each keyword is associated with a node of the graph structure.
Two nodes were connected if their keywords identify two similar mission specification problems. 
For example, the keywords ``visit'' and ``reach'' are related since in both cases the robot has to  visit/reach a location.

{\scshape{(Step.3)}} Since our interest  was not a mere classification of actions and movements that are executed by a robots, but rather detecting mission specification problems that concern how actions and movements are executed by a robot behavior over time,  nodes that contain keywords that only refer to actions are removed (e.g., balance).

{\scshape{(Step.4)}} Nodes that were connected through edges
 and contained keywords that identify to the same mission specification problem, e.g., visit and reach, were merged.

{\scshape{(Step.5)}} We organized the mission specification problems into a catalog represented through a  tree structure that facilitates browsing among mission specification problems.

The material produced in these steps can be found in our online appendix~\cite{paperstuff}.

\textbf{Pattern Formulation.}
We formulated patterns by following established practices in the literature~\cite{dwyer1999patterns,grunske2008specification,autili2015aligning}. 
A pattern is characterized by 
(i) a name;
(ii) a statement that captures the pattern intent (i.e., the mission requirement);
(iii) a template instance of the mission specification in LTL and CTL; 
(iv) variations describing possible minor changes that can be applied to the pattern;
(v) examples of known uses;
(vi) relationships of the pattern to others and;
(vii) occurrences of the pattern in literature.
For each LTL pattern we also designed a B\"uchi Automaton (BA) that unambiguously describes the behaviors of the system allowed by the mission specification.
The mission specification 
was designed by consulting specifications encoding requirements already present in the papers surveyed, by crosschecking them, and consulting  specification patterns already proposed in the software-engineering literature~\cite{autili2015aligning}.
If the proposed specification was related (or corresponded) with one of an already existing pattern, 
we indicated 
this in the relationships of the pattern to others, meaning that the pattern presented in literature is useful also to solve the identified mission specification problem.

\section{Mission Specification Patterns}
\label{sec:patterns}
\noindent
In this section, we present our catalog of mission specification patterns  and 
briefly present  one of them (\secref{sec:patterncatalog}).
We also present \psalmist, a tool that supports developers in systematic mission design.
 \psalmist\  supports the  description of mission requirements through the proposed patterns  and 
 the automatic generation of  mission specifications   
(\secref{sec:PsAlMISt}).

\pgfdeclarelayer{background}
\pgfdeclarelayer{foreground}

\newcommand{\coremovementcs}{4.8cm}

\begin{table*}
\caption{Core movement patterns}
\label{tab:coremovementpatterns}
\scriptsize
\begin{tabular}{ p{0.3cm}  p{4cm}  p{8cm} p{\coremovementcs} }
\toprule
 & \textbf{Description} & \textbf{Example} & \textbf{Formula} ($l_1, l_2, \ldots$ are location propositions) \\
\midrule
\multirow{1}{*}[-0.4ex]{\rotatebox[origin=c]{90}{\emph{Visit}}} & 
Visit a set of locations in an unspecified order.
& 
Locations $l_1$, $l_2$, and $l_3$ must be visited.
$l_1 \rightarrow l_4 \rightarrow l_3 \rightarrow l_1 \rightarrow l_4 \rightarrow l_2 \rightarrow (l_{\#})^\omega$ is an example trace that satisfies the mission requirement. & 
{\parbox[t]{\coremovementcs}{$\overset{n}{\underset{i=1}{\bigwedge}} \LTLf (l_i)$}} \\
\midrule
\raisebox{-0.7\normalbaselineskip}[0pt][0pt]{\rotatebox[origin=c]{90}{\parbox{1.5cm}{\emph{Sequenced\\ Visit}}}}
 & 
Visit a set of locations in sequence, one after the other. 
& 
Locations $l_1$, $l_2$, $l_3$ must be  covered following this sequence.
The trace  $l_1 \rightarrow l_4 \rightarrow l_3 \rightarrow l_1 \rightarrow l_4 \rightarrow l_2 \rightarrow (l_{\# \setminus 3})^\omega$  violates the mission since $l_3$ does not follow $l_2$.
The trace $l_1 \rightarrow l_3 \rightarrow l_1 \rightarrow l_2 \rightarrow l_4 \rightarrow l_3 \rightarrow (l_{\#})^\omega$ satisfies the mission requirement. & 
{\parbox[t]{\coremovementcs}{$\LTLf (l_1 \wedge \LTLf(l_2 \wedge \ldots \LTLf(l_n)))$}} \\
\midrule
\raisebox{-1.\normalbaselineskip}[0pt][0pt]{\rotatebox[origin=c]{90}{\parbox{1.5cm}{\emph{Ordered\\ Visit}}}} & 
Sequence visit does not forbid to visit  a successor location before its predecessor, but only that after the predecessor is visited the successor is also visited.
Ordered visit forbids a successor to be visited before its predecessor.
 & 
 Locations $l_1$, $l_2$, $l_3$ must covered following this order.
The trace $l_1 \rightarrow l_3 \rightarrow l_1 \rightarrow l_2 \rightarrow l_3 \rightarrow  (l_{\#})^\omega$ does not satisfy the mission requirement since $l_3$ preceeds $l_2$.
The trace $l_1 \rightarrow l_4 \rightarrow l_1 \rightarrow l_2 \rightarrow l_4 \rightarrow l_3 \rightarrow  (l_{\#})^\omega$ satisfies the  mission requirement. & 
{\parbox[t]{\coremovementcs}{$\LTLf (l_1 \wedge \LTLf(l_2 \wedge \ldots \LTLf(l_n))) $ \newline
$\overset{n-1}{\underset{i=1}{\bigwedge}} (\neg l_{i+1}) \LTLu l_i$}}
\\
\midrule
\raisebox{-2\normalbaselineskip}[0pt][0pt]{\rotatebox[origin=c]{90}{\parbox{1.5cm}{\emph{Strict Ordered\\ Visit}}}} &  
Ordered visit pattern does not avoid a predecessor location to be visited multiple times before its successor.
Strict ordered visit forbids this behavior. 
& 
Locations $l_1$, $l_2$, $l_3$ must be  covered following the strict order $l_1$, $l_2$, $l_3$.
The trace $l_1 \rightarrow l_4 \rightarrow l_1 \rightarrow l_2 \rightarrow l_4 \rightarrow l_3 \rightarrow (l_{\#})^\omega$ does not satisfy the  mission requirement since $l_1$ occurs twice before $l_2$.
The trace $l_1 \rightarrow l_4 \rightarrow l_2 \rightarrow l_4 \rightarrow l_3 \rightarrow (l_{\#})^\omega$ satisfies the mission requirement. & 
{\parbox[t]{\coremovementcs}{
$\LTLf (l_1 \wedge \LTLf(l_2 \wedge \ldots \LTLf(l_n))) $\\  
$\overset{n-1}{\underset{i=1}{\bigwedge}} (\neg l_{i+1}) \LTLu l_i$\\
$\overset{n-1}{\underset{i=1}{\bigwedge}} (\neg l_{i}) U (l_i \wedge \LTLx (\neg l_{i} \LTLu (l_{i+1})))$ }}
 \\
\midrule
\raisebox{-1.\normalbaselineskip}[0pt][0pt]{\rotatebox[origin=c]{90}{\parbox{1.5cm}{\emph{Fair\\ Visit}}}} & 
The difference among the number of times  locations within a set are visited is at most one. 
& 
Locations $l_1$, $l_2$, $l_3$ must be  covered in a fair way.
The trace $l_1 \rightarrow l_4 \rightarrow l_1 \rightarrow l_3 \rightarrow l_1 \rightarrow l_4 \rightarrow l_2 \rightarrow (l_{\# -\{1,2,3\}})^\omega$ does not perform a fair visit since it visits $l_1$ three times while $l_2$ and $l_3$ are visited once.
The trace $l_1 \rightarrow l_4 \rightarrow l_3 \rightarrow l_1 \rightarrow l_4 \rightarrow l_2 \rightarrow l_2 \rightarrow l_4 \rightarrow (l_{\# \setminus\{1,2,3\}})^\omega$  performs a fair visit since it visits locations $l_1$, $l_2$, and $l_3$ twice. & 
{\parbox[t]{\coremovementcs}{$\overset{n}{\underset{i=1}{\bigwedge}} \LTLf (l_i) $\\
$\overset{n}{\underset{i=1}{\bigwedge}} \LTLg (l_{i} \rightarrow \LTLx ((\neg l_i) \LTLw l_{(i+1)\%n}))$}}
\\
\midrule
\raisebox{-0.65\normalbaselineskip}[0pt][0pt]{\rotatebox[origin=c]{90}{\parbox{1.5cm}{\emph{Patrolling}}}} & 
Keep visiting a set of locations, but not in a particular  order. 
 &
Locations $l_1$, $l_2$, $l_3$ must be  surveilled.
 The trace $l_1 \rightarrow l_4 \rightarrow l_3 \rightarrow l_1 \rightarrow l_4 \rightarrow l_2  \rightarrow (l_2 \rightarrow l_3 \rightarrow l_1)^\omega$ ensures that  the mission requirement is satisfied.
 The trace $l_1 \rightarrow l_2 \rightarrow l _3  \rightarrow (l_1  \rightarrow l_3)^\omega$ represents a violation, since  $l_2$ is not surveilled.  & 
 {\parbox[t]{\coremovementcs}{$ \overset{n}{\underset{i=1}{\bigwedge}} \LTLg  \LTLf (l_i)$}} \\
 \midrule
\raisebox{-1\normalbaselineskip}[0pt][0pt]{\rotatebox[origin=c]{90}{\parbox{1.5cm}{\emph{Sequenced\\ Patrolling}}}} & 
Keep visiting a set of locations in sequence, one after the other.
 & 
 Locations $l_1$, $l_2$, $l_3$ must be patrolled in sequence. 
The trace $l_1 \rightarrow l_4 \rightarrow l_3\rightarrow l_1 \rightarrow l_4 \rightarrow l_2 \rightarrow (l_1 \rightarrow l_2 \rightarrow l_3)^\omega$ satisfies the mission requirement since globally any  $l_1$ will be  followed by $l_2$ and $l_2$ by $l_3$.
 The trace $l_1 \rightarrow l_4 \rightarrow l_3\rightarrow l_1 \rightarrow l_4 \rightarrow l_2 \rightarrow ( l_1 \rightarrow l_3)^\omega$ violates the mission requirement since  after visiting $l_1$, the robot does not visit $l_2$. & 
 {\parbox[t]{\coremovementcs}{$\LTLg (\LTLf (l_1 \wedge \LTLf(l_2 \wedge \ldots \LTLf(l_n))))$}} \\
 \midrule
\raisebox{-1\normalbaselineskip}[0pt][0pt]{\rotatebox[origin=c]{90}{\parbox{1.5cm}{\emph{Ordered\\ Patrolling}}}} & 
Sequence patrolling does not forbid to visit a successor location before its predecessor. 
Ordered patrolling ensures that (after a successor is visited) the successor is not visited (again) before its predecessor.
&
Locations $l_1$, $l_2$, and $l_3$ must be  patrolled  following the order $l_1$, $l_2$, and $l_3$.
 The trace $l_1\rightarrow l_4 \rightarrow l_3 \rightarrow l_1 \rightarrow l_4 \rightarrow l_2 \rightarrow ( l_1 \rightarrow l_2 \rightarrow l_3)^\omega$  violates the mission requirement since $l_3$ precedes $l_2$.
The trace $l_1 \rightarrow l_1 \rightarrow l_2 \rightarrow l_4 \rightarrow l_4 \rightarrow l_3 \rightarrow ( l_1 \rightarrow l_2 \rightarrow l_3)^\omega$  satisfies the mission requirement &  
{\parbox[t]{\coremovementcs}{$\LTLg (\LTLf (l_1 \wedge \LTLf(l_2 \wedge \ldots \LTLf(l_n)))) $\\
$\overset{n-1}{\underset{i=1}{\bigwedge}} (\neg l_{i+1}) \LTLu l_i$\\
$\overset{n}{\underset{i=1}{\bigwedge}} \LTLg (l_{(i+1)\%n} \rightarrow \LTLx ( (\neg l_{(i+1)\%n}) \LTLu l_{i}))$
}}\\
 \midrule
\raisebox{-2.6\normalbaselineskip}[0pt][0pt]{\rotatebox[origin=c]{90}{\parbox{1.5cm}{\emph{Strict Ordered\\ Patrolling}}}} &
Ordered patrolling pattern does not avoid a predecessor location to be visited multiple times before its successor.
Strict Ordered Patrolling ensures that, after a predecessor is visited, it is not visited again before its successor.
&
 Locations $l_1$, $l_2$, $l_3$ must be patrolled following the strict order $l_1$, $l_2$, and $l_3$.
The trace $l_1 \rightarrow l_4 \rightarrow l_1 \rightarrow l_2 \rightarrow l_4 \rightarrow l_3 \rightarrow ( l_1 \rightarrow l_2 \rightarrow l_3)^\omega$ violates the mission requirement since $l_1$ is visited twice before $l_2$.
The trace $l_1\rightarrow l_4 \rightarrow l_2 \rightarrow l_4 \rightarrow l_3 \rightarrow ( l_1 \rightarrow l_2 \rightarrow l_3)^\omega$ satisfies the mission requirement.
& 
{\parbox[t]{\coremovementcs}{$\LTLg (\LTLf (l_1 \wedge \LTLf(l_2 \wedge \ldots \LTLf(l_n)))) $ \\
$\overset{n-1}{\underset{i=1}{\bigwedge}} (\neg l_{i+1}) \LTLu l_i$\\
$\overset{n}{\underset{i=1}{\bigwedge}} \LTLg (l_{(i+1)\%n} \rightarrow \LTLx ( (\neg l_{(i+1)\%n}) \LTLu l_{i}))$ \\
$\overset{n-1}{\underset{i=1}{\bigwedge}} \LTLg ((l_{i}) \rightarrow \LTLx (\neg l_{i} \LTLu (l_{(i+1)\%n})))$ }}  \\  
\midrule
\raisebox{-1\normalbaselineskip}[0pt][0pt]{\rotatebox[origin=c]{90}{\parbox{1.5cm}{\emph{Fair\\ Patrolling}}}} & 
Keep visiting a set of locations and ensure that the difference among the number of times  locations within a set are visited is at most one. 
&
Locations $l_1$, $l_2$, and $l_3$ must be fair patrolled.
The trace $l_1 \rightarrow l_4 \rightarrow l_3 \rightarrow l_1 \rightarrow l_4 \rightarrow l_2 \rightarrow ( l_1 \rightarrow l_2 \rightarrow l_1 \rightarrow l_3)^\omega$ violates the mission requirements since the robot patrols $l_1$  more than  $l_2$ and $l_3$.
The trace $l_1 \rightarrow l_4 \rightarrow l_3 \rightarrow l_4 \rightarrow l_2  \rightarrow l_4 \rightarrow ( l_1 \rightarrow l_2 \rightarrow l_3)^\omega$ satisfies the mission requirement since locations  $l_1$, $l_2$, and $l_3$ are patrolled fairly.
 & 
{\parbox[t]{\coremovementcs}{ $\overset{n}{\underset{i=1}{\bigwedge}} \LTLg( \LTLf (l_i))$ \\
$\overset{n}{\underset{i=1}{\bigwedge}} \LTLg (l_{i} \rightarrow \LTLx ((\neg l_i) \LTLw l_{(i+1)\%n}))$}}
\\
 \bottomrule
\end{tabular}
\end{table*}

\newcommand{\triggercs}{5.8cm}

\begin{table*}
\caption{Avoidance and Trigger patterns.}
\label{tab:avoidanceAndTrigger}
\scriptsize
\begin{tabular}{ p{0.6cm}  p{2.5cm}  p{7.8cm}  p{\triggercs} }
\toprule
 & \textbf{Description} & \textbf{Example} & \textbf{Formula} \\
\midrule 
\hspace{-0.2cm}\parbox[t]{0.6cm}{\emph{Past\\ avoidance}}
& 
A condition has been fulfilled in the past.
 &
If the robot enters location $l_1$, then it should have not visited location $l_2$ before.
The trace $l_3 \rightarrow l_4 \rightarrow l_1 \rightarrow l_2 \rightarrow l_4 \rightarrow l_3 \rightarrow (  l_2 \rightarrow l_3)^\omega$ satisfies the mission requirement since location $l_2$ is not entered before location $l_1$.
 & {\parbox[t]{\triggercs}{$(\neg (l_1)) \LTLu p$, where $l_1 \in L$ and $p \in M$}}  \\
 \midrule
\hspace{-0.2cm}\parbox[t]{0.6cm}{\emph{Global\\ avoidance}}
 & 
An avoidance condition globally holds throughout the mission. &  
The robot should avoid entering location $l_1$.
Trace $l_3 \rightarrow l_4 \rightarrow l_3 \rightarrow l_2 \rightarrow l_4 \rightarrow l_3 \rightarrow ( l_3 \rightarrow l_2 \rightarrow l_3)^\omega$, satisfies the mission requirement since the robot never enters 
$l_1$.  & 
{\parbox[t]{\triggercs}{$\LTLg(\neg (l_1))$, where $l_1 \in L$}} \\
\midrule 
\hspace{-0.2cm}\parbox[t]{0.6cm}{\emph{Future\\ avoidance}}
  &
 After the occurrence of an event, avoidance has to be fulfilled. & 
If the robot enters  $l_1$, then it should  avoid entering $l_2$ in the future.
The trace $l_3 \rightarrow l_4 \rightarrow l_3 \rightarrow l_1 \rightarrow l_4 \rightarrow l_3 \rightarrow ( l_3 \rightarrow l_2 \rightarrow l_3)^\omega$  does not satisfy the mission requirement since  $l_2$ is entered after  $l_1$. & 
{\parbox[t]{\triggercs}{$\LTLg( (c) \rightarrow( \LTLg (\neg  (l_1))))$,  where $c \in M$  and $l_1 \in PL$}} \\
\midrule 
\hspace{-0.2cm}\parbox[t]{0.6cm}{\emph{Upper Rest.\\ Avoidance}}
 & 
A restriction on the maximum number of occurrences is desired. & 
A robot has to visit  $l_1$ at most $3$ times.
The trace $l_1 \rightarrow l_4 \rightarrow l_1 \rightarrow l_3 \rightarrow l_1 \rightarrow l_4 \rightarrow l_1 \rightarrow ( l_3)^\omega$ violates the mission requirement since $l_1$ is visited four times.
The trace $l_4 \rightarrow l_3 \rightarrow l_1 \rightarrow l_2 \rightarrow l_4 \rightarrow ( l_3)^\omega$ satisfies the mission requirement. & 
{\parbox[t]{\triggercs}{$\neg \LTLf (\underbrace{l_1 \wedge \LTLx (\LTLf(l_1 \wedge \ldots \LTLx (\LTLf(l_1)}_\text{n})))) $, where $l_1 \in L$
}} \\
\midrule
\hspace{-0.2cm}\parbox[t]{0.6cm}{\emph{Lower Rest.\\ Avoidance}}
 & 
A restriction on the minimum number of occurrences is desired. &
A robot to enter location $l_1$ at least $3$ times.
The trace  $l_4 \rightarrow l_3 \rightarrow l_2 \rightarrow l_2\rightarrow  l_4 \rightarrow  ( l_3)^\omega$ violates the mission requirement since location $1$ is never entered. 
The trace $l_1 \rightarrow l_4 \rightarrow l_3 \rightarrow l_1 \rightarrow l_4 \rightarrow l_1 \rightarrow  ( l_3)^\omega$ satisfies the mission requirement. & 
{\parbox[t]{\triggercs}{$\LTLf (\underbrace{l_1 \wedge \LTLx (\LTLf(l_1 \wedge \ldots \LTLx (\LTLf(l_1)}_\text{n})))) $, where $l_1 \in L$}}\\
\midrule
\hspace{-0.2cm}\parbox[t]{0.6cm}{\emph{Exact Rest.\\ Avoidance}} & 
The number of occurrences desired is an exact number. &
A robot must enter location $l_1$ exactly $3$ times.
The trace  $l_4 \rightarrow l_3 \rightarrow l_2 \rightarrow l_2 \rightarrow l_4 \rightarrow  ( l_3)^\omega$ violates the mission requirement. 
The trace $l_1 \rightarrow l_4 \rightarrow l_3 \rightarrow l_1 \rightarrow l_4 \rightarrow l_1 \rightarrow  ( l_3)^\omega$ satisfies the mission requirement since  location $l_1$ is entered exactly $3$ times. & 
{\parbox[t]{\triggercs}{$\underbrace{(\neg (l1)) \LTLu (l1 \wedge
(\LTLx ((\neg  l1) \LTLu (l1
\ldots 
\wedge (\LTLx ((\neg  l1) \LTLu (l1}_\text{n}$ $
\wedge (\LTLx (\LTLg (\neg  l1))))))))))$, where $l_1 \in L$}} \\
\midrule
\hspace{-0.2cm}\parbox[t]{0.6cm}{\emph{Inst.\\ Reaction}}
& 
The occurrence of a stimulus instantaneously triggers a counteraction. &
When location $l_2$ is reached action $a$ must be executed.
The trace $l_1 \rightarrow l_3 \rightarrow \{ l_2,a \} \rightarrow \{ l_2,a \} \rightarrow l_4 \rightarrow (l_3)^\omega$ satisfies  the mission requirement since when location $l_2$ is entered condition $a$ is performed. 
The trace $l_1 \rightarrow l_3 \rightarrow l_2 \rightarrow \{l_1,a\} \rightarrow l_4 \rightarrow (l_3)^\omega$ does not satisfy the mission requirement since when $l_2$ is reached  $a$ is  not executed. & $\LTLg( p_1 \rightarrow p_2)$, where $p_1 \in M$   and $p_2 \in PL \cup PA$ \\
\midrule
\hspace{-0.2cm}\parbox[t]{0.6cm}{\emph{Delayed\\ Reaction}} & 
The occurrence of a stimulus triggers a counteraction some time later & 
When $c$ occurs the robot must start moving toward location $l_1$, and $l_1$ is subsequently finally reached.
The trace $l_1 \rightarrow l_3 \rightarrow \{l_2,c\} \rightarrow l_1 \rightarrow l_4 \rightarrow (l_3)^\omega$ satisfies the mission requirement, since after $c$ occurs the robot starts moving toward location $l_1$, and location $l_1$ is finally reached.
The trace $l_1 \rightarrow l_1 \rightarrow \{ l_2, c\} \rightarrow l_3 \rightarrow (l_3)^\omega$
does not satisfy the mission requirement since $c$ occurs when the robot is in  $l_2$, and  $l_1$ is not finally reached.
& 
{\parbox[t]{\triggercs}{$\LTLg( p_1 \rightarrow \LTLf (p_2))$, where $p_1 \in M$  and $p_2 \in PL \cup PA$}}\\
\midrule
\hspace{-0.2cm}\parbox[t]{0.6cm}{\emph{Prompt\\ Reaction}} &  
The occurrence of a stimulus triggers a counteraction promptly, i.e. in the next time instant.
& If $c$ occurs  $l_1$ is reached in the next time instant.
The trace $l_1 \rightarrow l_3 \rightarrow \{l_2,c\} \rightarrow l_1 \rightarrow l_4 \rightarrow (l_3)^\omega$ satisfies the mission requirement, since after $c$ occurs   $l_1$  is reached within the next time instant.
The trace $l_1 \rightarrow$  $l_3 \rightarrow \{l_2,c\} \rightarrow l_4 \rightarrow l_1 \rightarrow (l_3)^\omega$ does not satisfy the mission requirement.
& 
{\parbox[t]{\triggercs}{$\LTLg( p_1 \rightarrow \LTLx (p_2))$, where $p_1 \in M$  and $p_2 \in PL \cup PA$}}\\
\midrule
\hspace{-0.2cm}\parbox[t]{0.6cm}{\emph{Bound\\ Reaction}} &
A counteraction must be performed  every time and only when a specific location is entered. & 
Action $a_1$ is bound though a delay to location $l_1$.
The trace $l_1 \rightarrow l_3 \rightarrow \{l_2,c\} \rightarrow \{l_1,a_1\} \rightarrow l_4 \rightarrow \{l_1,a_1\} \rightarrow (l_3)^\omega$ satisfies the mission requirement.
The trace $l_1 \rightarrow l_3 \rightarrow \{l_2,c\} \rightarrow \{l_1,a_1\} \rightarrow \{l_4,a_1\} \rightarrow \{l_1,a_1\} \rightarrow (l_3)^\omega$ does not satisfy the mission requirement since $a_1$ is executed in location $l_4$. & 
{\parbox[t]{\triggercs}{$\LTLg( p_1 \leftrightarrow p_2)$,  where $p_1 \in M$  and $p_2 \in PL \cup PA$}} \\
\midrule
\hspace{-0.2cm}\parbox[t]{0.6cm}{\emph{Bound\\ Delay}}
 & 
 A counteraction must be performed, in the next time instant,  every time and only when a specific location is entered &
Action $a_1$ is bound to location $l_1$.
The trace $l_1 \rightarrow l_3 \rightarrow \{l_2,c\} \rightarrow \{l_1\} \rightarrow \{l_4,1_1\} \rightarrow \{l_1\} \rightarrow \{l_4,a_1\} \rightarrow (l_3)^\omega$ satisfies the mission requirement.
The trace $l_1 \rightarrow l_3 \rightarrow \{l_2,c\} \rightarrow \{l_1\} \rightarrow \{l_4,1_1\} \rightarrow \{l_1,a_1\} \rightarrow \{l_4\} \rightarrow (l_3)^\omega$ does not satisfy the mission requirement. & 
{\parbox[t]{\triggercs}{$\LTLg( p_1 \leftrightarrow \LTLx (p_2))$,  where $p_1 \in M$  and $p_2 \in PL \cup PA$}}   \\
\midrule
\hspace{-0.2cm}\parbox[t]{0.6cm}{\emph{Wait}}
& 
Inaction is desired till a stimulus occurs. 
& 
The robot  remains in location $l_1$ until  condition $c$ is satisfied. 
The trace $l_1 \rightarrow l_3 \rightarrow \{ l_2,c\} \rightarrow l_1 \rightarrow l_4 \rightarrow (l_3)^\omega$ violates the mission requirement since the robot left $l_1$ before condition $c$ is satisfied.
The trace $l_1 \rightarrow \{l_1,c\} \rightarrow l_2 \rightarrow l_1 \rightarrow l_4 \rightarrow (l_3)^\omega$ satisfies the mission requirement.
&  $(l_1) \LTLu (p) $,  where $l_1 \in L$  and $p \in PE \cup PA$ \\
\bottomrule
\end{tabular}
\vspace{-0.5cm}
\end{table*}

\subsection{Mission Specification Pattern Catalog}
\label{sec:patterncatalog}
\noindent
Our catalog of robotic mission specification patterns comprises $22$ patterns
organized into a pattern tree as illustrated in \figref{fig:specificationPatternSystem}.
Leaves of the tree 
represent mission specification patterns. 
Intermediate nodes facilitate browsing within the hierarchy and aid pattern selection and decision making.
Patterns  identified by following the procedure described in Sec.~\ref{sec:methodology} are graphically indicated with a solid border.

Due to space limits, we provide a high-level description of all patterns identified, examples of application,  and the corresponding LTL mission specifications.
The interested reader may refer to our online appendix~\cite{paperstuff}, which contains additional examples, occurrences of patterns in the literature,   
relations among the patterns and additional CTL mission specifications.

\pattern{Strict Ordered Patrolling}{
A robot must patrol a set of locations following a strict sequence ordering.
Such locations can be, e.g., areas in a building 
to be surveyed.
}{
The following formula encodes the mission in LTL for $n$ locations and a robot $r$ (\% is the modulo arithmetic operator):
\begin{center}
$\overset{n}{\underset{i=1}{\bigwedge}} \LTLg (\LTLf (l_1 \wedge \LTLf (l_2 \wedge ... \LTLf (l_n)))) 
\overset{n-1}{\underset{i=1}{\bigwedge}}
 ((\neg l_{i+1})\ U\ l_i) 
 \overset{n}{\underset{i=1}{\bigwedge}} 
 \LTLg (l_{(i+1)\%n} \rightarrow \LTLx((\neg l_{(i+1)\%n})\ U\ l_i)) $\\
\end{center}
Example with two locations.
\begin{center}
$\LTLg (\LTLf (l_1 \wedge \LTLf (l_2))) \wedge ((\neg l_2)\ U\ l_1) \wedge \LTLg (l_2 \rightarrow \LTLx((\neg l_2)\ U\ l_1)) \wedge \LTLg(l_1 \rightarrow \LTLx((\neg l_1)\ U\ l_2))$\\
\end{center}
where 
$l_1$ and $l_2$ are expressions that indicate that a robot $r$ is in locations $l_1$ and $l_2$, respectively.  }
{
A developer may want to allow traces in which sequences of \emph{consecutive} $l_1$ ($l_2$) are allowed, that is strict ordering is applied on sequences of non consecutive $l_1$ ($l_2$). In this case, traces in the form $l_1 \rightarrow (\rightarrow l_1 \rightarrow l_1 \rightarrow l_3 \rightarrow l_2)^\omega$ are admitted, while traces in the form $l_1 \rightarrow (\rightarrow l_1 \rightarrow l_3 \rightarrow l_1 \rightarrow l_2)^\omega$ are not admitted.  This variation can be encoded using the following specification:
\begin{center}
$\LTLg (\LTLf (l_1 \wedge \LTLf (l_2))) \wedge ((\neg l_2)\ U\ l_1) \wedge \LTLg ((l_2 \wedge \LTLx(\neg l_2)) \rightarrow \LTLx((\neg l_2)\ U\ l_1)) \wedge \LTLg ((l_1 \wedge \LTLx(\neg l_1)) \rightarrow \LTLx((\neg l_1)\ U\ l_2))$
\end{center}
This specification allows for sequences of consecutive $l_1$ ($l_2$) since the left side of the implication $l_1 \wedge \LTLx(\neg l_1)$ ($l_2 \wedge \LTLx(\neg l_2)$) is only triggered when  $l_1$ ($l_2$) is exited.
}{
A common usage example of the Strict Ordered Patrolling pattern is a scenario where a robot is performing surveillance in a building during night hours.
Strict Sequence Patrolling and Avoidance often go together.
Avoidance patterns are used to force robots to avoid obstacles as they guard a location.
Triggers can also be used in combination with the Strict Sequence Patrolling pattern to specify conditions upon which Patrolling should start or stop. 
}{
The Strict Ordered Patrolling pattern is a specialisation of the Ordered Patrolling  pattern, forcing the strict  ordering.
}{
 Smith et. al.~\cite{smith2011optimal} proposed a mission specification  forcing a robot to not visit a location twice in a row before a  target location is reached. 
}{strictorderedpatrolling.png}{width=0.7\textwidth} 

\textbf{Preliminaries.} 
To aid comprehension of behavior and facilitate precise pattern definitions, we introduce the following notation.
Given a finite set of locations $L=\{ l_1, l_2, \ldots, l_n\}$ and robots $R=\{ r_1, r_2, \ldots, r_n\}$,
$PL=\{ r_x\ in\ l_y \mid  r_x \in R \text{ and } l_y \in L\}$ is a set of location propositions, each indicating that a robot $r_x$ is in a specific location $l_y$ of the environment. 
Given a finite set of  conditions of the environment $C=\{ c_1, c_2, \ldots, c_m \}$, we indicate as $PE=\{s_1, s_2, \ldots, s_m \}$ a set of propositions such that  $s_i \in PE$ is true if and only if condition $c_i$ holds.
Given a finite set of  actions that the robots can perform $A=\{ a_1, a_2, \ldots, a_m \}$, we indicate as $PA=\{ r_x\ exec\ a_y \mid r_x \in R \text{ and } a_y \in A \}$ a set of propositions such that  $r_x\ exec\ a_y$ is true if and only if action $a_y$ is performed by robot $r_x$.
We define the  set of propositions  $M$ of a robotic application as $PL \cup PE \cup PA$.
A trace is an infinite sequence  $M_x \rightarrow M_y \rightarrow M_z \ldots$ where $M_x,M_y,M_z \subseteq M$ indicate a trace in which  $M_z$ holds after $M_y$, and $M_y$ holds after $M_x$.
For example,  $\{r_1\ in\ l_1 \} \rightarrow \{r_1\ in\ l_2, c_1 \} \rightarrow \{ c_2, r_2\ exec\ a_1 \} \ldots$  is  a trace where the element in position $1$ of the trace indicates that the robot $r_1$ is in location $l_1$, then the  element in position $2$ indicates that the robot $r_1$ is in location $l_2$ and condition $c_1$ holds (indicating, for example, that an obstacle is detected), and then the  element in position $3$ indicates that condition $c_2$ holds and robot  $r_2$ is executing action $a_1$.
In the following, with a slight abuse of notation, when a set is a singleton we will omit brackets.
We use the notation $(M_x \rightarrow \ldots \rightarrow M_y )^\omega$, where $M_x,\ldots ,M_y \subseteq M$,  to indicate a sequence $M_x \rightarrow \ldots \rightarrow M_y $ that occurs infinitely.
We use the notation $l_{\#}$ to indicate any location, e.g., $r_1\ in\ l_1\rightarrow r_1\ in\ l_{\#} \rightarrow r_1\ in\ l_2$ indicates that a robot $r_1$ visits location $l_1$, afterwards any  location, and then location $l_2$.
We use the notation $l_{\# \setminus K}$, where 
$K \subset M$,  to indicate any possible location not in $K$, e.g., 
$r_1\ in\ l_1\rightarrow r_1\ in\ l_{\# \setminus \{l_3\}} \rightarrow r_1\ in\ l_2$ indicates that $r_1$ visits $l_1$, then any location except $l_3$ is visited, and finally $l_2$.

\textbf{Patterns.} Patterns are organized in three main groups -- core movement (Table~\ref{tab:coremovementpatterns}), triggers (Table~\ref{tab:avoidanceAndTrigger}), and avoidance (Table~\ref{tab:avoidanceAndTrigger}), explained in the following.
For simplicity, in Tables~\ref{tab:coremovementpatterns} and~\ref{tab:avoidanceAndTrigger}, we assume that a single robot is considered during the mission specification and
we use the notation $l_x$ as shortcut for $r_1\ in\ l_x$.
The examples assume that the environment is made by four locations, namely  $l_1$, $l_2$, $l_3$, and $l_4$.

\emph{Core movement patterns}. How robots should move within an environment can be divided in two major categories representing locations' coverage and locations' surveillance. Coverage patterns require a robot to reach a set of locations of the environment. 
Surveillance patterns require a robot to  \emph{keep} reaching a set of  locations  of the environment.

\emph{Avoidance patterns.} Robot movements may be constrained in order to avoid occurrence of some behavior (Table~\ref{tab:avoidanceAndTrigger}).
Avoidance may reflect a condition or be a bound to the occurrence of some event.
\emph{Conditional avoidance} generally holds globally (i.e., for the entire behavior) and applies when avoidance of locations or obstacles is sought that depends on some condition.
For example, a cleaning robot may avoid visiting locations that have been already cleaned.
In the \emph{restricted avoidance} case, avoidance does not hold globally but accounts for a number of occurrences of an avoidance case.
Depending on the number of occurrences being a maximum, minimum  or exact number, 
\emph{upper}, \emph{exact} or \emph{lower} restricted avoidance is yielded.
For example, a cleaning robot may avoid cleaning a room more than three times. 

\emph{Trigger patterns.} Robot's reactive behaviour based on stimuli, or robot's inaction until a stimulus occurs are expressed as trigger patterns in Table~\ref{tab:avoidanceAndTrigger}.  

As an example, the definition of the \emph{Strict Ordered Patrolling} mission specification pattern is presented in Fig.~\ref{fig:Strict Ordered Patrolling}. The patterns in detail are available in our online appendix~\cite{paperstuff}.

\subsection{Specification Pattern Tool Support}
\label{sec:PsAlMISt}
\begin{figure}[t]
\centering
	\includegraphics[width=1\columnwidth]{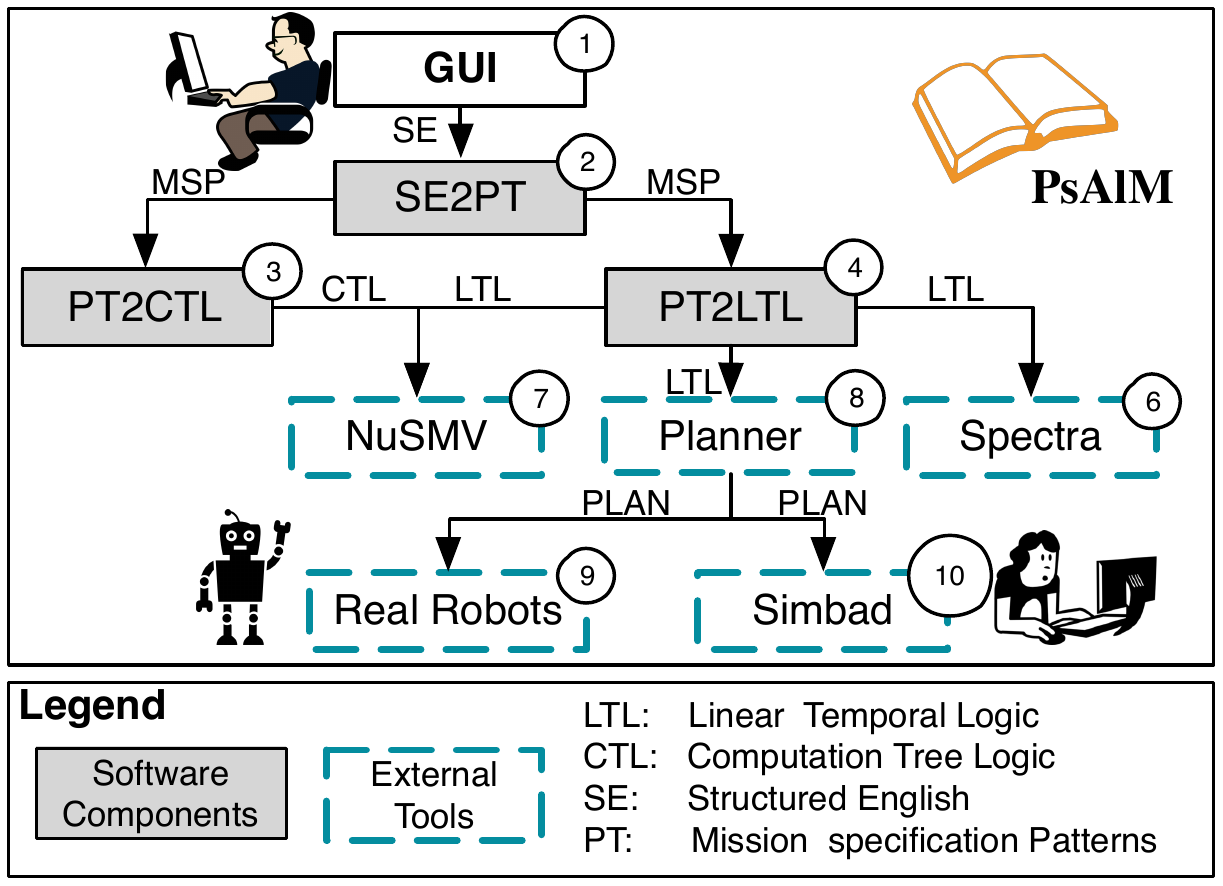}
	\caption{Main components of the \psalmist tool}
	\label{fig:psaimist}
\end{figure}

\noindent
We used the proposed pattern catalog to express robotic missions requirements and to automatically generate their mission specifications.
To support developers in mission design we implemented the tool \psalmist ~\cite{PSALM}, which allows creating complex mission requirements by composing patterns with simple operators.
\psalmist transforms mission requirements (i.e., composed patterns) into mission specifications in LTL or CTL.
\Figref{fig:psaimist} illustrates the components of \psalmist.

\psalmist provides a GUI \circled{\scriptsize{1}} that allows the definition of robotic missions requirements
 through a structured English grammar, which uses patterns as basic building blocks and AND and OR logic operators to compose these patterns. 
  The structured English grammar and the \psalmist tool are provided in our online appendix~\cite{paperstuff}.
The \texttt{SE2PT} component extracts from a  mission requirement the set of patterns that are composed through the AND and OR  operators  \circled{\scriptsize{2}}.
The \texttt{PT2LTL} \circled{\scriptsize{3}} and \texttt{PT2CTL} \circled{\scriptsize{4}} components automatically generate LTL and CTL specifications from these patterns.

The produced LTL specifications  
can be used in different ways --
three possible usages are presented in \figref{fig:psaimist}.
The LTL formulae are 
(i) fed into an existing planner and used to generate plans that satisfy the mission specification \circled{\scriptsize{5}};
(ii) converted into Deterministic B\"uchi automata  used as input to the widely used Spectra~\cite{Spectra} robotic application modeling tool \circled{\scriptsize{6}}; and
(iii) converted into the NuSMV~\cite{cimatti1999nusmv} input language to be used as input for model checking \circled{\scriptsize{7}}.
The plans produced using the planner are (i) 
used as inputs by the Simbad~\cite{hugues2006simbad} simulation package  \circled{\scriptsize{10}}, which is an autonomous robot simulation package for education and research; and
(ii) performed by actual real robots \circled{\scriptsize{9}}, as also illustrated in the following section. 
Produced CTL specifications are also converted into the NuSMV~\cite{cimatti1999nusmv} input language to be used as input for model checking 
\circled{\scriptsize{7}}.

\section{Evaluation}
\label{sec:results}
\noindent
Our evaluation addressed the following two questions. \textbf{RQ1:} How \emph{effective} is  the pattern catalog in capturing mission requirements and producing mission specifications? \textbf{RQ2:} Are the proposed mission specifications  \emph{correct}? 

\begin{table*}[t]
\scriptsize
\caption{Mission specification patterns for \emph{Exp1}. Labels SC1, SC2, \ldots SC5 identify the considered scenarios.}
\begin{tabular}{ c  p{14cm}  p{2.4cm} } 
\toprule
\multicolumn{1}{  c }{\textbf{SC}} & \multicolumn{1}{  c }{\textbf{Description}} & \multicolumn{1}{  c }{\textbf{Patterns}} \\
\midrule
SC1 & A robot is deployed within a supermarket and reports about the absence of sold items within a set of locations (i.e. $l1$, $l2$, $l3$, and $l4$). 
Furthermore, if in location $l4$ (where water supplies are present) a human is detected, it has to perform a collaborative grasping action and help the human in placing new water supplies. & 
Ordered Patrolling,\newline Instantaneous Reaction
\\
\midrule
SC2 & Three robots are deployed within an hospital environment: a mobile platform (Summit~\cite{Summit}), a manipulator (PA10~\cite{PA10}) and a mobile manipulator (Tiago~\cite{Tiago}), identified in the following as MP, M and MM, respectively.
The robot M is deployed in hospital storage; when  items (e.g., towels) are needed by a nurse or doctors, M has to load them on the MP. MP should reach the location where the nurse is located. 
If the item is heavy (e.g., heavy medical equipment), MM should reach the location where the nurse is to help unloading the equipment. 
When MP and MM are not required for shipping items they are patrolling a set of locations to avoid unauthorized people entering restricted areas of the hospital (e.g., radiotherapy rooms). & 
Patrolling,\newline Instantaneous Reaction,\newline Ordered Visit,\newline Wait\\
\midrule
SC3 & A robot is developed within a university building to deliver coffee to employees.
The robot reaches the coffee machine, uses the coffee machine to prepare the coffee and delivers it to the employee. & 
Strict Ordered Visit,\newline Instantaneous Reaction\\
\midrule
SC4 & A robot is deployed within a shop to check the presence of intruders during night time. 
It has to iteratively check for intruders and report on their presence 
& 
Patrolling,\newline Instantaneous Reaction\\
\midrule
SC5 & A robot is deployed within a company to notify employees in presence of a fire alarm. 
If a fire is detected, the robot is send to different areas of the company to ask employees to leave the building. &
Visit,\newline Instantaneous Reaction\\
\bottomrule
\end{tabular}
\label{fig:exp1}
\end{table*}

\textbf{Coverage of Real-World Missions (RQ1).}
\noindent
We investigated (i) how the pattern catalog supports the specification of mission requirements  and (ii) how the pattern catalog reduces ambiguities in mission requirements. 

\emph{Exp1.} We checked how the pattern catalog supports the formulation of mission requirements (and the generation of mission specifications) in real-world robotic scenarios. 
To this end, we defined five scenarios (\tabref{fig:exp1}) in collaboration with our industrial partners (\BOSCH\ and \PAL).

\begin{table}[t]
\centering
\caption{Results of experiment Exp2.  
Lines contain the total number of mission requirements (MR), the number of not expressible (NE) and ambiguous (A)  mission requirements and
the number of cases that lead to a consensus (C) and no consensus (NC).}
\scriptsize
\begin{tabular}{ l@{\hspace{1.5mm}}  l@{\hspace{1.5mm}}  l@{\hspace{1.5mm}}  l@{\hspace{1.5mm}}  l@{\hspace{1.5mm}}  l@{\hspace{1.5mm}}  l@{\hspace{1.5mm}}  l@{\hspace{1.5mm}}  l@{\hspace{1.5mm}}   l@{\hspace{1.5mm}}  l@{\hspace{1.5mm}}  l@{\hspace{1.5mm}}  l@{\hspace{1.5mm}}  l@{\hspace{1.5mm}} } 
\toprule
& \multicolumn{11}{  c }{Spectra Robotic Application} & & \\
\midrule
 & 1 & 2 & 3 & 4 & 5 & 6 & 7 & 8 & 9 & 10 & \multicolumn{1}{c }{11} & \multicolumn{1}{ c }{MP}  & \multicolumn{1}{ c }{\textbf{Total}}\\
\midrule
MR & 29 & 2 & 22 & 5 & 1 & 159 & 4  & 32  & 47 & 53 & \multicolumn{1}{c }{74} & \multicolumn{1}{ c }{8}  & \multicolumn{1}{ c }{436} \\
 NE  & 3 & 0 & 0 & 0 & 0 & 47 & 0 & 0 & 7 & 1 & \multicolumn{1}{c }{8} & \multicolumn{1}{ c }{0} & \multicolumn{1}{ c }{66}\\
 A &  3 & 0 & 2 & 1 & 0 & 35 & 0 & 10 & 12 & 32 & \multicolumn{1}{c }{7} & \multicolumn{1}{ c }{0} & \multicolumn{1}{ c }{102}\\
  C &   13 & 0 & 11 & 2 & 1 & 29 & 4 & 8 & 11 & 8 & \multicolumn{1}{c }{20} & \multicolumn{1}{ c }{5} & \multicolumn{1}{ c }{112}\\
   NC & 10 & 2 & 9 & 2 & 0 & 48 & 0 & 14 & 17 & 12 & \multicolumn{1}{c }{39} & \multicolumn{1}{ c }{3} & \multicolumn{1}{ c }{156}\\
\bottomrule
\end{tabular}
\label{tab:exp2missionrequirements}
\end{table}

The pattern catalog supported the creation of mission requirements using the patterns listed in \tabref{fig:exp1} for the different scenarios.
In all the scenarios, \psalmist allowed the automatic creation of LTL mission specifications from the mission requirements without any human intervention. 
The mission specifications were then executed by the robots by relying on existing planners (see \figref{fig:psaimist}). 
Videos of the robots performing the described missions 
are available in our dedicated website~\cite{paperstuff}. 
The pattern catalog effectively supports the creation of mission requirements and specifications in realistic, industry-sourced scenarios.

\emph{Exp2.} We collected mission requirements in natural language from available requirements produced from Spectra~\cite{Spectra} and LTLMoP~\cite{finucane2010ltlmop,wei2016extended}. 
Spectra is a tool that supports the design of the robotic applications. 
LTLMoP is a software package 
designed to assist in the development, implementation, and testing of high-level robot 
controllers.
We checked  how the pattern catalog may have supported developers in the definition of the mission requirements. 

In the case of Spectra, we used the Spectra files to extract mission requirements for robotic systems. 
In total, 11 robotic applications were considered. 
Note that mission requirements are realistic since they were finally executed with real robots~\cite{Examples}.
We automatically  extracted \numbermissionrequirementspectra\ mission requirements from the Spectra file.
The number of mission requirements (MR) per robotic application is reported in \tabref{tab:exp2missionrequirements}.
In the case of LTLMoP, $8$ requirements were extracted from  the corresponding research papers~\cite{finucane2010ltlmop,wei2016extended} (\tabref{tab:exp2missionrequirements} MP column). 

\begin{table}
\caption{Results of experiment Exp2. Number of occurrences of each pattern in the considered mission requirements.}
\scriptsize
\begin{tabular}{  l@{\hspace{1.5mm}} l@{\hspace{1.5mm}}  l@{\hspace{1.5mm}} l@{\hspace{1.5mm}} l@{\hspace{1.5mm}} l@{\hspace{1.5mm}}  l@{\hspace{1.5mm}} l@{\hspace{1.5mm}}  l@{\hspace{1.5mm}}  l@{\hspace{1.5mm}} l@{\hspace{1.5mm}}  l@{\hspace{1.5mm}} }
\toprule
Pattern & Occ & Pattern & Occ  & Pattern & Occ & Pattern & Occ & Pattern & Occ  \\
\midrule
Visit  & 25  & SeqVisit &  1 & OrdVisit &  1 & InstReact & 127 &  GlobAvoid & 25\\
   PastAvoid & 60 & DelReact & 50 & Wait & 3 & FutAvoid & 48 & SeqVisit & 1  \\
    StrictOrdPat & 1 & OrdVisit & 1 & ExactRest & 1 && \\
\bottomrule
\end{tabular}
\label{tab:exp2}
\end{table}

Each mission requirement was independently analyzed by two of the authors. 
The authors checked whether it is possible to express the mission requirement using the mission specification patterns. 
If one  the authors stated that the requirement is not expressible the requirement is marked as  not expressible (NE).
The  number of not expressible mission requirements is presented in Table~\ref{tab:exp2} under the column with header NE.
If at least one of the authors found the mission requirement is ambiguous she marked it with the flag $A$.
Otherwise, the mission requirement is labeled with the mission specification patterns needed to express the mission requirement. 
Then, the mission specification patterns used to express the mission requirement are considered.
If the authors used the same mission specification patterns to express the mission requirement, a consensus is reached.
The number of  mission requirements that leads to consensus (resp. no consensus) is indicated in the row labeled C (resp. NC).
The number of occurrences of each pattern is indicated in Table~\ref{tab:exp2}.

The results show that most of the mission requirements ($370$ over $436$) were expressible using the pattern catalog, which is a reasonable coverage for pattern catalog usage.
The $66$ mission requirements that are not covered suggested the introduction of new patterns identified in Fig.~\ref{fig:specificationPatternSystem} with a dashed border.
It also shows that the pattern catalog is effective in real case scenarios. 
In $102$ cases the mission requirements were ambiguous, meaning that different interpretations can be given to the proposed mission requirement.
In these cases, alternative combinations of patterns have been proposed by the authors to express the mission requirement.
Each of these alternatives represents a possible way of expressing it in a non-ambiguous manner.
In $156$ cases, while the authors judged that the requirement was not ambiguous, different pattern combinations were proposed.
The combinations of patterns encode possible ways of expressing the mission requirement in  a non-ambiguous manner.

\emph{Exp3.} We analyzed the mission specifications contained in the Spectra examples collected in Exp2.
We collected \missionNocompilationError\ distinct LTL mission specifications 
and we analyzed each of these specifications\footnote{This number differs from the one of Exp2, since some specifications were not related with a mission requirement in the form of natural language.}.
We verified whether it is possible to obtain the mission specifications starting from the proposed 
patterns, by performing the following steps.

{\scshape{(Step.1)}} For each property we automatically checked whether it was an instance of a mission specification pattern or a simple combination of mission specification patterns. Results are shown in Table~\ref{tab:exp3}.
Among \missionNocompilationError\ mission specifications \steponematching\ were obtainable from the proposed patterns. 

{\scshape{(Step.2)}} We considered the properties that did not match any of the proposed patterns. 
$127$ of these properties are simple statements on the initial state of the system (no temporal operator is used), and thus did not match  any of the proposed patterns. 
$442$ formulae concern properties that refer to variation of the trigger patterns that we have added to the pattern catalogue.
$224$ formulae still did not match any of the proposed patterns.
After analysis, $155$ among them were expressed using past temporal operators, which are not used in the  mission specifications proposed 
in this work. In step 3 we checked whether these specifications might be reformulated without the past operators.
 $69$ of these properties, while they can be rewritten using the proposed patterns, they are written as complex LTL formulae and thus they do not match any of our patterns or combination of them.
 
 {\scshape{(Step.3)}} We considered the $155$ properties expressed using past temporal operators and we designed mission specifications for them.
 We found that $129$ of the proposed LTL formulae  match one of the proposed pattern,  while $26$ are complex LTL formulae that did not match any of the patterns.  
 Thus, the final coverage of the proposed pattern catalog is $92\%$.

 We then analyzed $13$ mission specifications expressed in the form of LTL properties considered in~\cite{ruchkin2018ipl} and $22$ PCTL properties considered in~\cite{ruchkin2018ipl}, transformed  in CTL by replacing the probabilistic operator ($\mathcal{P}$) with the universal  quantifier ($\forall$).
 Given the small number of mission specifications we manually checked the presence of patterns in the  formulae (Step 1 in Table~\ref{tab:exp3}).
  The results show that the pattern system was able to  generate almost all  mission specifications  ($1154$ over $1251$).

\begin{table}
\centering
\scriptsize
\caption{Results of experiment Exp3. Pattern occurrence in the considered mission specifications.}
\begin{tabular}{ c  c  c  c  c }
\toprule
&  & \multicolumn{2}{c }{LTL}  & CTL \\ 
\midrule
& \multicolumn{1}{c }{Pattern} & Spectra & \cite{ruchkin2018ipl} & \cite{ruchkin2018ipl} \\
\midrule
\multirow{5}{*}{\raisebox{\normalbaselineskip}[0pt][0pt]{\rotatebox[origin=c]{90}{\parbox{2cm}{Step 1}}}}& \multicolumn{1}{c }{Instantaneous reaction} & 318 & 0 &  0\\
 & \multicolumn{1}{c }{Visit} & 52 & 0 & 0\\
  & \multicolumn{1}{c }{Patrolling} & 0 & 1 & 0\\
   & \multicolumn{1}{c }{Strict Ordered Visit} & 0 & 9 &18 \\
  & \multicolumn{1}{c }{Wait} & 0 & 1 & 2\\
 & \multicolumn{1}{c }{Avoidance/Invariant} & 21 & 0 & 0 \\
 & \multicolumn{1}{c }{Visit and Instantaneous reaction} & 18 & 0 & 0\\
  & \multicolumn{1}{c}{Strict Ordered Visit and Global Avoidance} & 0 & 0 & 1\\ 
 & \multicolumn{1}{c}{Reaction chain (chain of instantaneous reactions)} & 15 & 0 & 0 \\
 & \multicolumn{1}{c}{Non matching} & 792  & 1 & 1 \\
\midrule
\multirow{6}{*}{\raisebox{\normalbaselineskip}[0pt][0pt]{\rotatebox[origin=c]{90}{\parbox{1cm}{ Step 2}}}} & \multicolumn{1}{c }{Init} & 127 & - & -\\
 & \multicolumn{1}{c }{Fast reaction} & 379 & - & -\\
 & \multicolumn{1}{c }{Binded reaction} & 36 & - & -\\
 & \multicolumn{1}{c }{Binded delay} & 27 & - & -\\
 & \multicolumn{1}{c }{Non matching for past} & 155 & - & -\\
 & \multicolumn{1}{c }{Actual non matching} & 69  & - & - \\
\midrule
\multirow{3}{*}{\raisebox{\normalbaselineskip}[0pt][0pt]{\rotatebox[origin=c]{90}{\parbox{1cm}{Step 3}}}}  & \multicolumn{1}{c}{Fast reaction} & 103 & - & - \\
 & \multicolumn{1}{c }{Binded delay} & 26 & - & - \\
& \multicolumn{1}{c }{Actual non matching} & 26 & - & -\\
\bottomrule
\end{tabular}
\label{tab:exp3}
\end{table}

\begin{table}
\centering
\scriptsize
\caption{Results  of experiments \emph{Exp4}, \emph{Exp5} and \emph{Exp6}.
For \emph{Exp4} columns contain the number of times a plan is found ($\top$)  and not found ($\bot$).
For \emph{Exp5} and \emph{Exp6} columns contain the number of times the mission requirement is satisfied ($\top$)  and violated ($\bot$).}
\label{tab:expRQ2}
\begin{tabular}{ c  c  c  c  c  c c }
\toprule
&   \multicolumn{2}{c }{Exp4}  & \multicolumn{2}{c }{Exp5}  & \multicolumn{2}{c }{Exp6}  \\ 
\midrule
\multicolumn{1}{ c }{\textbf{Mission Requirement}} & $\top$ & $\bot$ & $\top$ & $\bot$ & $\top$ & $\bot$ \\
\midrule
\multicolumn{1}{ c }{OrdPatrol,UpperRestAvoid,Wait} & 2 & 10   & 1 & 11 & 1 & 11 \\
\multicolumn{1}{ c }{FairVisit,ExactRestAvoid$^\ast$,DelReact} & 5 & 7 & 0 & 12 & 4 & 8 \\
\multicolumn{1}{ c }{StrOrdVisit,GlobalAvoid,InstReact} & 3 & 9 & 1 & 11 & 1 & 11\\
\multicolumn{1}{ c }{SeqVisit,FutAvoid,BindDel$^\ast$} & 1 & 11 & 0 & 12 & 2 & 10\\
\multicolumn{1}{ c }{OrdVisit,PastAvoid,InstReact} & 3 & 9 & 1 & 11 & 1 & 11\\
\multicolumn{1}{ c }{Visit,LowRestAvoid,BindReact} & 3 & 9 & 1 & 11 & 1 & 11\\
\multicolumn{1}{ c }{StrictOrdPatrol,FutAvoid,Wait} & 1 & 11 & 1 & 11 & 1 & 11\\
\multicolumn{1}{ c }{Patrol,LowRestAvoid,InstReact}&  3 & 9  & 1 & 11 & 1 & 11\\
\multicolumn{1}{ c }{FairPatrol,ExactRestAvoid$^\ast$,DelReact} & 3 & 9 & 0 & 12 & 4 & 8\\
\multicolumn{1}{ c }{SeqPatrol,UpperRestAvoid,FastReact$^\ast$} & 1 & 11 & 0 & 12 & 2 & 10\\
\bottomrule
\end{tabular}
\end{table}

\textbf{Summary.}  
The pattern catalog is effective  in supporting developers in defining mission requirements and in 
generating mission specifications.
Exp1 and Exp2 show that the pattern catalog effectively supports the definition of mission requirements. 
and that helps in reducing ambiguities in   available mission requirements. 
Exp1 and Exp3 show that the pattern catalog effectively supports the generation of mission specifications.
Exp1 shows how the pattern catalog can be used to generate precise, unambiguous, and formal mission specifications in industry sourced  scenarios.

\textbf{Correctness of the Patterns (RQ2).}
\noindent
To verify the mission specifications (LTL and CTL formulas) 
we manually reviewed them and performed a random testing to confirm that the specifications do
not permit undesired system behaviors that were not detected during the manual check.

\emph{Manual check.} We manually inspected  instances of the patterns obtained by fixing the number of locations to be visited, conditions to be considered etc. 
For LTL formulae we used SPOT~\cite{spot} to generate B\"uchi automata (BA) encoding the traces of the system allowed and forbidden by the specification. 
We manually inspected the BA of all the proposed patterns.

\emph{Random testing.} 
We performed some testing by exploiting a set of randomly generated models: a widespread technique to evaluate artifacts in the software engineering community~\cite{tabakov2005experimental,de2006antichains,tabakov2007model,saadatpanah2012comparing,famelis2012partial,FM2016,fase2018}, also used in the robotic community~\cite{menghi2018multi,DBLP:conf/icse/MenghiGPT18,best2016multi,takacs2009multi,stentz1996map}.
We  generated  $12$ scenarios 
representing the structure of buildings containing $16$ locations, where a robot is deployed.
The building has been generated by allocating $12$ traversable locations  and $4$ locations that cannot be crossed, on a $4 \times 4$ matrix.
Identifiers  $l_0, l_1, \ldots, l_{11}$ are randomly assigned to the traversable locations.
In  $6$ of the $12$ scenarios 
the robot can move among adjacent cells that are traversable, while it cannot move within not crossable locations.
In the other $6$ scenarios 
the robot can cross the adjacent cells by respecting the following rules: (i) it can move from a traversable cell with coordinate $[i,j]$  to a traversable cell with coordinate $[i,j+1]$ and $[i+1,j]$;
(ii) it can move from a traversable cell with coordinate $[i,j]$ to another with coordinate $[h,k]$, where $i$ (resp. $h$) is the maximum (resp. minimum) row index of a cell that corresponds to a traversable location and $h$ (resp. $k$) is the maximum (resp. minimum) column index of the traversalble locations at row $i$ (resp. $h$).
Conditions and actions are treated by considering whether  a box is present in a location ($cond$ in the following), and  the capability of the robot in changing its color ($act$ in the following).
We randomly select $4$ traversable locations in which $cond$ is true and $4$ locations in which $act$ can be performed.

For each scenario we considered different mission requirements; each 
obtained by randomly combining 
a core movement, a trigger  and  an avoidance pattern, and by ensuring that each pattern is used in at least one mission requirement.
In total we generated $10$ mission requirements (Table~\ref{tab:expRQ2}). 
Core movement patterns are parametrized with 
locations $l_1, l_2$.
The upper, exact and lower restricted avoidance patterns are parametrized 
by forcing the robot to visit location $l_3$, at most, exactly, and at least $2$ times, respectively.
The global avoidance  pattern forces the robot to not visit $l_3$, while the future and past avoidance force the robot to not visit $l_3$ after and before condition $cond$ is satisfied, i.e., a room that contains a box is visited.
The wait pattern forces the robot to wait in  location $l_4$  if a box is not present.
The other trigger patterns are parametrized with the action $act$ that must be executed by the robot in relation with the occurrence of condition $cond$.
We subsequently performed the following experiments.

\emph{Exp4.} We generated the LTL specifications of the considered mission requirements. We (i) negated the LTL specification; (ii) encoded the specification and the model of the scenario in NuSMV~\cite{cimatti1999nusmv}; (iii) used NuSMV to check whether the models contained a path that satisfied the mission specification (violates its negation). 
If a plan was present we used Simbad~\cite{hugues2006simbad} to simulate the robot executing the plan.
We verified whether the results were correct:  when we expected a plan to not be present in the given model, NuSMV was not able to compute it, and, 
when a plan was expected to be present it was computed by NuSMV. We also checked the correctness of the generated plans using the Simbad simulator.
Results confirm the correctness of the  LTL mission specifications.
The column labeled with the $\top$ (resp. $\bot$) symbol of Table~\ref{tab:expRQ2} contains the number of cases in which a plan was (resp. was not) present.

\emph{Exp5.} We generated LTL and CTL specifications for the considered mission requirements. We (i) encoded the LTL and CTL specifications and the model of the scenario in NuSMV~\cite{cimatti1999nusmv}; (iii) we used NuSMV to check whether the verification of the specifications returned the same results.
Table~\ref{tab:expRQ2} contains the number of cases in which the mission requirement was satisfied  ($\top$) and not satisfied ($\bot$).
Mission requirements were generally not satisfied, since for being satisfied they have to hold on all the paths of the models.
NuSMV always returned the same results for LTL and CTL specifications confirming the correctness of 
CTL  specifications. 

\emph{Exp6.} We investigated 
why in several cases the mission requirement was always not satisfied.
In these cases we  relaxed the mission requirements, by removing 
the patterns marked with the $^\ast$ symbol in Table~\ref{tab:expRQ2}. 
We executed the same steps of \emph{Exp4}.
Table~\ref{tab:expRQ2} confirmed that by relaxing the mission requirements there were cases in which the mission requirement was actually satisfied. 
This is a further confirmation  that the mission specifications are correct.

\section{Discussion and Related Work}
\label{sec:discussion}

\noindent We discuss the proposed patterns and present related work.

\emph{Methodology.} 
The number of mission requirements analyzed 
is in line with other approaches in the field~\cite{dwyer1999patterns,grunske2008specification,konrad2005real,autili2015aligning,bianculli2012specification}.
These requirements usually come from exemplar scenarios 
used to provide evaluation about effectiveness of research-intensive works.
As such, we believe that the scope of the pattern system is quite wide.
Our study is certainly not exhaustive, as 
(i) formal specification in robotic application spreads, and
(ii) the types of mission specifications change over time.
As shown in the evaluation,  
patterns will grow over time 
as specifications that do not belong to the 
catalog are provided. 

\emph{Patterns.} While the presented patterns 
are mainly conceived to address needs of robotic mission specification, 
they are more generic and can be applied when the need is to specify some ``ordering" among events or action execution. 
Rather than predicate on robots reaching a set of actions, coverage and surveillance patterns  may  also include propositions that refer to generic events.
In this sense,  the proposed patterns can be considered as an extension of the property specification patterns~\cite{OrderSpecificationPatterns,dwyer1999patterns} that explicitly address different ordering among the occurrence of a set of events.
While in this paper we proposed a direct encoding in LTL and CTL, they may also be expressed in terms of standard  property specification patterns. 
The instantaneous reaction pattern may be obtained from the response pattern scoped with the global operator.
The precedence chain and the response chains~\cite{OrderSpecificationPatterns,dwyer1999patterns} (that illustrate the 2 cause-1 effect and 1 cause-2 effects chain), can be composed with the precedence and response patterns to specify different ordering among a set of events.

\emph{Evaluation.} The Spectra tool 
only supports specifications captured by the GR(1) LTL fragment used to describe three types of  guarantees: initial, safety, and liveness.
Initial guarantees constraint the initial states of the environment. 
Safety guarantees start with the temporal operator $\LTLg$ and constraints the current and next state.
Liveness guarantees start with the temporal operators $\LTLg \LTLf$ and may not include the $\LTLx$ operator.
These constraints justify the prevalence of patterns presented in Tables~\ref{tab:exp2},~\ref{tab:exp3}, and~\ref{tab:expRQ2}.
While the proposed patterns can be expressed using deterministic B\"uchi automata (DBA), which can be translated in GR(1) formulae~\cite{maozsynthesis}, 
a manual encoding of the proposed patterns in GR(1) is complex and error prone.
This is confirmed by the fact that analysis on the standard property specification patterns that can be expressed in GR(1), and an automatic procedure to map these patterns on formulae that are in the GR(1) fragment has been recently conducted~\cite{maozsynthesis}.
All of the patterns proposed in this work are expressible using GR(1) formulae, and the automatic procedure presented in~\cite{maozsynthesis} can be integrated in \psalmist\ to generate Spectra formulae.

\textbf{Related work.} 
Temporal logic specification patterns  are a well-known solution to support developers in requirement specification~\cite{dwyer1999patterns,konrad2005real,grunske2008specification,autili2015aligning,Paun99,Remenska2014,Castillos2013}.
Property specification patterns use in specific domains have  been investigated in literature, including service-based applications~\cite{bianculli2012specification}, safety~\cite{Bitsch2001} and security~\cite{Spanoudakis2007}.
However, at the best of our knowledge, no work has considered mission patterns for robotic applications.

Domain Specific Languages (DSLs)~\cite{Schmidt2006,Ciccozzi4496,Ruscio2016,Bozhinoski2015,Adam2014} have been 
proposed for various purposed including production and analysis of behaviour descriptions, property verification and planning.
However, features incorporated within DSLs are usually arbitrarily chosen by relying on the domain-specific experience of robotic engineers. Instead,  specification patterns presented in this paper are collected from missions encountered in scientific literature, evaluated in industrial uses, and aim at supporting a wide range of robotic needs.
We believe that the presented patterns consist of basic building blocks that can be reused within existing and new robotic DSLs. 
Moreover, support for developers on solving the mission specification problem is also provided in literature by graphical tools that simplify the specification of LTL formulae~\cite{lee1997graphical,smith2001events,srinivas2013graphical}. 
Our work is  complementary with those; 
graphical logic mission specifications can also be integrated within \psalmist .

\section{Conclusion}
\label{sec:conclusion}

\noindent We proposed a pattern catalog for mission specification of mobile robots. 
We identified patterns by analyzing mission requirements that have been systematically collected from scientific publications. 
We  presented \psalmist , a tool that uses the proposed patterns  to support developers in designing complex missions.
We evaluated (ii) the support provided by the catalog in the definition of real-world missions; (ii) the correctness of the mission specifications contained in our pattern catalog.

Future extensions of our mission specification pattern catalog will consider also time, space, and probability. 
We will also investigate 
the use of spatial logics~\cite{aiello2007handbook,papadimitriou1996topological,bivand2013spatial,erwtimatiko,cardelli2002spatial} to express more complex spatial robotic behaviours and perform user studies.

\balance
\bibliographystyle{IEEEtran}  

\bibliography{sample-bibliography-short} 

\begin{thebibliography}{100}
\providecommand{\url}[1]{#1}
\csname url@samestyle\endcsname
\providecommand{\newblock}{\relax}
\providecommand{\bibinfo}[2]{#2}
\providecommand{\BIBentrySTDinterwordspacing}{\spaceskip=0pt\relax}
\providecommand{\BIBentryALTinterwordstretchfactor}{4}
\providecommand{\BIBentryALTinterwordspacing}{\spaceskip=\fontdimen2\font plus
\BIBentryALTinterwordstretchfactor\fontdimen3\font minus
  \fontdimen4\font\relax}
\providecommand{\BIBforeignlanguage}[2]{{%
\expandafter\ifx\csname l@#1\endcsname\relax
\typeout{** WARNING: IEEEtran.bst: No hyphenation pattern has been}%
\typeout{** loaded for the language `#1'. Using the pattern for}%
\typeout{** the default language instead.}%
\else
\language=\csname l@#1\endcsname
\fi
#2}}
\providecommand{\BIBdecl}{\relax}
\BIBdecl

\bibitem{wrs:online}
{IFR}, ``{World Robotic Survey},'' {\footnotesize
  \url{https://ifr.org/ifr-press-releases/news/world-robotics-survey-service-robots-are-conquering-the-world-}},
  2016.

\bibitem{brugali2007software}
D.~Brugali, \emph{Software engineering for experimental robotics}.\hskip 1em
  plus 0.5em minus 0.4em\relax Springer, 2007, vol.~30.

\bibitem{Lee2008}
E.~A. Lee, ``Cyber physical systems: Design challenges,'' in \emph{Symposium on
  Object Oriented Real-Time Distributed Computing (ISORC)}.\hskip 1em plus
  0.5em minus 0.4em\relax IEEE, 2008, pp. 363--369.

\bibitem{Perez2008}
J.~P{\'e}rez, N.~Ali, J.~A. Cars\i, I.~Ramos, B.~\'{A}lvarez, P.~Sanchez, and
  J.~A. Pastor, ``Integrating aspects in software architectures: Prisma applied
  to robotic tele-operated systems,'' \emph{Information and Software
  Technology}, vol.~50, no. 9-10, pp. 969--990, 2008.

\bibitem{Gamez2013563}
N.~Gamez and L.~Fuentes, ``Architectural evolution of famiware using
  cardinality-based feature models,'' \emph{Information and Software
  Technology}, vol.~55, no.~3, pp. 563--580, 2013.

\bibitem{4799437}
D.~Brugali and E.~Prassler, ``Software engineering for robotics,'' \emph{IEEE
  Robotics Automation Magazine}, vol.~16, no.~1, pp. 9--15, March 2009.

\bibitem{Gotz:2018:RIW:3149485.3149523}
S.~G\"{o}tz, C.~Piechnick, and A.~Wortmann, ``Report on the 4th international
  workshop on model-driven robot software engineering ({MORSE}),'' \emph{ACM
  SIGSOFT Software Engineering Notes}, 2018.

\bibitem{luckcuck2018formal}
M.~Luckcuck, M.~Farrell, L.~Dennis, C.~Dixon, and M.~Fisher, ``Formal
  specification and verification of autonomous robotic systems: A survey,''
  \emph{arXiv preprint arXiv:1807.00048}, 2018.

\bibitem{Maoz:2018:SEC:3196558.3196561}
S.~Maoz and J.~O. Ringert, ``On the software engineering challenges of applying
  reactive synthesis to robotics,'' in \emph{Workshop on Robotics Software
  Engineering}, ser. RoSE '18.\hskip 1em plus 0.5em minus 0.4em\relax ACM,
  2018.

\bibitem{MaozFSE}
S.~{M}aoz and J.~O. Ringert, ``On well-separation of {GR}(1) specifications,''
  in \emph{Foundations of Software Engineering (FSE)}.\hskip 1em plus 0.5em
  minus 0.4em\relax ACM, 2016.

\bibitem{lignos2015provably}
C.~Lignos, V.~Raman, C.~Finucane, M.~Marcus, and H.~Kress-Gazit, ``Provably
  correct reactive control from natural language,'' \emph{Autonomous Robots},
  vol.~38, no.~1, pp. 89--105, 2015.

\bibitem{Mindstorms}
http://www.lego.com/eng/education/mindstorms/default.asp, ``Lego.com
  educational division, mindstorms for schools.'' 2018.

\bibitem{Choregraph}
Softbankrobotics. Choregraph.
  \url{http://doc.aldebaran.com/1-14/dev/tools/robot-simulation.html}.
  Accessed: 2018-06-20.

\bibitem{DSLInRobotics}
A.~Nordmann, N.~Hochgeschwender, and S.~Wrede, ``A survey on domain-specific
  languages in robotics,'' in \emph{Simulation, Modeling, and Programming for
  Autonomous Robots}.\hskip 1em plus 0.5em minus 0.4em\relax Springer, 2014.

\bibitem{arkin2006missionlab}
R.~Arkin, ``Missionlab v7. 0,'' 2006.

\bibitem{Teambots}
\BIBentryALTinterwordspacing
T.~Balch, ``Teambots,'' 2004. [Online]. Available: \url{www.teambots.org}
\BIBentrySTDinterwordspacing

\bibitem{maoz2011aspectltl}
S.~Maoz and Y.~Sa'ar, ``Aspectltl: an aspect language for ltl specifications,''
  in \emph{International conference on Aspect-oriented software
  development}.\hskip 1em plus 0.5em minus 0.4em\relax ACM, 2011.

\bibitem{Ruscio2016}
D.~D. Ruscio, I.~Malavolta, P.~Pelliccione, and M.~Tivoli, ``Automatic
  generation of detailed flight plans from high-level mission descriptions,''
  in \emph{Model Driven Engineering Languages and Systems}, ser. MODELS.\hskip
  1em plus 0.5em minus 0.4em\relax ACM, 2016.

\bibitem{Broy:1991:DSD:952786.952788}
M.~Broy, ``Declarative specification and declarative programming,'' in
  \emph{Software Specification and Design}.\hskip 1em plus 0.5em minus
  0.4em\relax IEEE, 1991.

\bibitem{menghi2018multi}
C.~Menghi, S.~Garcia, P.~Pelliccione, and J.~Tumova, ``Multi-robot {LTL}
  planning under uncertainty,'' in \emph{International Symposium on Formal
  Methods (FM)}.\hskip 1em plus 0.5em minus 0.4em\relax Springer, 2018.

\bibitem{ulusoy2011optimal}
A.~Ulusoy, S.~L. Smith, X.~C. Ding, C.~Belta, and D.~Rus, ``Optimal multi-robot
  path planning with temporal logic constraints,'' in \emph{International
  Conference on Intelligent Robots and Systems (IROS)}.\hskip 1em plus 0.5em
  minus 0.4em\relax IEEE, 2011.

\bibitem{fainekos2009temporal}
G.~E. Fainekos, A.~Girard, H.~Kress-Gazit, and G.~J. Pappas, ``Temporal logic
  motion planning for dynamic robots,'' \emph{Automatica}, vol.~45, no.~2, pp.
  343--352, 2009.

\bibitem{guo2013revising}
M.~Guo, K.~H. Johansson, and D.~V. Dimarogonas, ``Revising motion planning
  under linear temporal logic specifications in partially known workspaces,''
  in \emph{International Conference on Robotics and Automation (ICRA)}.\hskip
  1em plus 0.5em minus 0.4em\relax IEEE, 2013.

\bibitem{wolff2013automaton}
E.~M. Wolff, U.~Topcu, and R.~M. Murray, ``Automaton-guided controller
  synthesis for nonlinear systems with temporal logic,'' in \emph{International
  Conference on Intelligent Robots and Systems (IROS)}.\hskip 1em plus 0.5em
  minus 0.4em\relax IEEE, 2013.

\bibitem{kress2011robot}
H.~Kress-Gazit, ``Robot challenges: Toward development of verication and
  synthesis techniques [errata],'' \emph{IEEE Robotics \& Automation Magazine},
  vol.~18, no.~4, pp. 108--109, 2011.

\bibitem{doi:10.1177/0278364914546174}
M.~Guo and D.~V. Dimarogonas, ``Multi-agent plan reconfiguration under local
  {LTL} specifications,'' \emph{The International Journal of Robotics
  Research}, 2015.

\bibitem{DBLP:journals/corr/MaozR16}
S.~Maoz and J.~O. Ringert, ``Synthesizing a lego forklift controller in
  {GR(1):} {A} case study,'' in \emph{Proceedings Fourth Workshop on Synthesis
  ({SYNT})}, 2015.

\bibitem{maozsynthesis}
S.~{M}aoz and J.~O. Ringert, ``{GR}(1) synthesis for {LTL} specification
  patterns,'' in \emph{Foundations of Software Engineering (FSE)}.\hskip 1em
  plus 0.5em minus 0.4em\relax ACM, 2015.

\bibitem{situa}
F.~S. Rodriguez, B.~C. Diego, V.~M. Rodilla, J.~Rodriguez-Aragon, R.~A. Santos,
  and C.~Fernandez-Carames, ``The complete integration of missionlab and
  carmen,'' \emph{International Journal of Advanced Robotic Systems}, vol.~14,
  no.~3, p. 1729881417703565, 2017.

\bibitem{hinchey2005requirements}
M.~G. Hinchey, J.~L. Rash, and C.~A. Rouff, ``Requirements to design to code:
  Towards a fully formal approach to automatic code generation,'' 2005.

\bibitem{Kramer2007DevelopmentEF}
J.~F. Kramer and M.~Scheutz, ``Development environments for autonomous mobile
  robots: A survey,'' \emph{Autonomous Robots}, vol.~22, pp. 101--132, 2007.

\bibitem{6907489m}
S.~Maniatopoulos, M.~Blair, C.~Finucane, and H.~Kress-Gazit, ``Open-world
  mission specification for reactive robots,'' in \emph{International
  Conference on Robotics and Automation (ICRA)}.\hskip 1em plus 0.5em minus
  0.4em\relax IEEE, 2014.

\bibitem{Bozhinoski2015}
D.~Bozhinoski, D.~D. Ruscio, I.~Malavolta, P.~Pelliccione, and M.~Tivoli,
  ``Flyaq: Enabling non-expert users to specify and generate missions of
  autonomous multicopters,'' in \emph{Automated Software Engineering
  (ASE)}.\hskip 1em plus 0.5em minus 0.4em\relax IEEE, 2015.

\bibitem{ding2011automatic}
X.~C. Ding, M.~Kloetzer, Y.~Chen, and C.~Belta, ``Automatic deployment of
  robotic teams,'' \emph{IEEE Robotics \& Automation Magazine}, vol.~18, no.~3,
  pp. 75--86, 2011.

\bibitem{lee1997graphical}
I.~Lee and O.~Sokolsky, ``A graphical property specification language,'' in
  \emph{High-Assurance Systems Engineering Workshop}.\hskip 1em plus 0.5em
  minus 0.4em\relax IEEE, 1997.

\bibitem{smith2001events}
M.~H. Smith, G.~J. Holzmann, and K.~Etessami, ``Events and constraints: A
  graphical editor for capturing logic requirements of programs,'' in
  \emph{International Symposium on Requirements Engineering}.\hskip 1em plus
  0.5em minus 0.4em\relax IEEE, 2001.

\bibitem{srinivas2013graphical}
S.~Srinivas, R.~Kermani, K.~Kim, Y.~Kobayashi, and G.~Fainekos, ``A graphical
  language for ltl motion and mission planning,'' in \emph{International
  Conference on Robotics and Biomimetics (ROBIO)}.\hskip 1em plus 0.5em minus
  0.4em\relax IEEE, 2013.

\bibitem{raman2013sorry}
V.~Raman, C.~Lignos, C.~Finucane, K.~C. Lee, M.~Marcus, and H.~Kress-Gazit,
  ``Sorry dave, i'm afraid i can't do that: Explaining unachievable robot tasks
  using natural language,'' University of Pennsylvania Philadelphia United
  States, Tech. Rep., 2013.

\bibitem{finucane2010ltlmop}
C.~Finucane, G.~Jing, and H.~Kress-Gazit, ``{LTLM}o{P}: Experimenting with
  language, temporal logic and robot control,'' in \emph{International
  Conference on Intelligent Robots and Systems (IROS)}.\hskip 1em plus 0.5em
  minus 0.4em\relax IEEE, 2010, pp. 1988--1993.

\bibitem{6016586}
X.~C. Ding, M.~Kloetzer, Y.~Chen, and C.~Belta, ``Automatic deployment of
  robotic teams,'' \emph{IEEE Robotics Automation Magazine}, vol.~18, no.~3,
  pp. 75--86, Sept 2011.

\bibitem{endo2004usability}
Y.~Endo, D.~C. MacKenzie, and R.~C. Arkin, ``Usability evaluation of high-level
  user assistance for robot mission specification,'' \emph{Transactions on
  Systems, Man, and Cybernetics, Part C (Applications and Reviews)}, vol.~34,
  no.~2, pp. 168--180, 2004.

\bibitem{wei2016extended}
W.~Wei, K.~Kim, and G.~Fainekos, ``Extended {LTL}vis motion planning
  interface,'' in \emph{International Conference on Systems, Man, and
  Cybernetics}.\hskip 1em plus 0.5em minus 0.4em\relax IEEE, 2016.

\bibitem{Lignos2015}
C.~Lignos, V.~Raman, C.~Finucane, M.~Marcus, and H.~Kress-Gazit, ``Provably
  correct reactive control from natural language,'' \emph{Autonomous Robots},
  vol.~38, no.~1, pp. 89--105, Jan 2015.

\bibitem{shah2015resolving}
U.~S. Shah and D.~C. Jinwala, ``Resolving ambiguities in natural language
  software requirements: a comprehensive survey,'' \emph{ACM SIGSOFT Software
  Engineering Notes}, 2015.

\bibitem{kiyavitskaya2008requirements}
N.~Kiyavitskaya, N.~Zeni, L.~Mich, and D.~M. Berry, ``Requirements for tools
  for ambiguity identification and measurement in natural language requirements
  specifications,'' \emph{Requirements engineering}, 2008.

\bibitem{ringert2014requirements}
J.~O. Ringert, B.~Rumpe, and A.~Wortmann, ``A requirements modeling language
  for the component behavior of cyber physical robotics systems,'' \emph{arXiv
  preprint arXiv:1409.0394}, 2014.

\bibitem{kress2009temporal}
H.~Kress-Gazit, G.~E. Fainekos, and G.~J. Pappas, ``Temporal-logic-based
  reactive mission and motion planning,'' \emph{Transactions on robotics},
  vol.~25, no.~6, pp. 1370--1381, 2009.

\bibitem{yoo2016online}
C.~Yoo, R.~Fitch, and S.~Sukkarieh, ``Online task planning and control for
  fuel-constrained aerial robots in wind fields,'' \emph{The International
  Journal of Robotics Research}, vol.~35, no.~5, pp. 438--453, 2016.

\bibitem{dwyer1999patterns}
M.~B. Dwyer, G.~S. Avrunin, and J.~C. Corbett, ``Patterns in property
  specifications for finite-state verification,'' in \emph{International
  Conference on Software Engineering (ICSE)}.\hskip 1em plus 0.5em minus
  0.4em\relax IEEE, 1999.

\bibitem{EMERSON1990995}
E.~A. EMERSON, ``\{CHAPTER\} 16 - temporal and modal logic,'' in \emph{Formal
  Models and Semantics}, ser. Handbook of Theoretical Computer Science, J.~V.
  LEEUWEN, Ed.\hskip 1em plus 0.5em minus 0.4em\relax Elsevier, 1990, pp. 995
  -- 1072.

\bibitem{5238617}
H.~Kress-Gazit, G.~E. Fainekos, and G.~J. Pappas, ``Temporal-logic-based
  reactive mission and motion planning,'' \emph{Transactions on Robotics},
  vol.~25, no.~6, pp. 1370--1381, 2009.

\bibitem{Holzmann2002}
G.~J. Holzmann, ``The logic of bugs,'' in \emph{Foundations of Software
  Engineering (FSE)}.\hskip 1em plus 0.5em minus 0.4em\relax ACM, 2002.

\bibitem{Autili2007}
M.~Autili, P.~Inverardi, and P.~Pelliccione, ``Graphical scenarios for
  specifying temporal properties: An automated approach,'' \emph{Automated
  Software Engg.}, vol.~14, no.~3, 2007.

\bibitem{grunske2008specification}
L.~Grunske, ``Specification patterns for probabilistic quality properties,'' in
  \emph{International Conference on Software Engineering (ICSE)}.\hskip 1em
  plus 0.5em minus 0.4em\relax IEEE, 2008.

\bibitem{konrad2005real}
S.~Konrad and B.~H. Cheng, ``Real-time specification patterns,'' in
  \emph{International conference on Software engineering (ICSE)}.\hskip 1em
  plus 0.5em minus 0.4em\relax IEEE, 2005.

\bibitem{autili2015aligning}
M.~Autili, L.~Grunske, M.~Lumpe, P.~Pelliccione, and A.~Tang, ``Aligning
  qualitative, real-time, and probabilistic property specification patterns
  using a structured english grammar,'' \emph{Transactions on Software
  Engineering}, vol.~41, no.~7, pp. 620--638, 2015.

\bibitem{bianculli2012specification}
D.~Bianculli, C.~Ghezzi, C.~Pautasso, and P.~Senti, ``Specification patterns
  from research to industry: a case study in service-based applications,'' in
  \emph{International Conference on Software Engineering (ICSE)}.\hskip 1em
  plus 0.5em minus 0.4em\relax IEEE, 2012.

\bibitem{Menghi:Idea}
C.~Menghi, C.~Tsigkanos, T.~Berger, P.~Pelliccione, and C.~Ghezzi, ``Property
  specification patterns for robotic missions,'' in \emph{International
  Conference on Software Engineering (ICSE): Companion Proceeedings}, 2018.

\bibitem{brooks1991intelligence}
R.~A. Brooks \emph{et~al.}, ``Intelligence without reason,'' \emph{Artificial
  intelligence: critical concepts}, vol.~3, pp. 107--63, 1991.

\bibitem{brugali2005software}
D.~Brugali and M.~Reggiani, ``Software stability in the robotics domain: issues
  and challenges,'' in \emph{International Conference on Information Reuse and
  Integration}.\hskip 1em plus 0.5em minus 0.4em\relax IEEE, 2005.

\bibitem{brugali2007stable}
D.~Brugali, ``Stable analysis patterns for robot mobility,'' in \emph{Software
  Engineering for Experimental Robotics}.\hskip 1em plus 0.5em minus
  0.4em\relax Springer, 2007, pp. 9--30.

\bibitem{PSALM}
C.~Menghi, C.~Tsigkanos, T.~Berger, and P.~Pelliccione, ``{PsAlM}:
  Specification of dependable robotic missions,'' in \emph{International
  Conference on Software Engineering (ICSE): Companion Proceeedings}, 2019.

\bibitem{Spectra}
S.~Maoz and J.~O. Ringert. Spectra.
  \url{http://smlab.cs.tau.ac.il/syntech/spectra/}. Accessed: 2018-06-20.

\bibitem{cimatti1999nusmv}
A.~Cimatti, E.~Clarke, F.~Giunchiglia, and M.~Roveri, ``{NuSMV}: A new symbolic
  model verifier,'' in \emph{Computer Aided Verification (CAV)}.\hskip 1em plus
  0.5em minus 0.4em\relax Springer, 1999.

\bibitem{hugues2006simbad}
L.~Hugues and N.~Bredeche, ``Simbad: an autonomous robot simulation package for
  education and research,'' in \emph{International Conference on Simulation of
  Adaptive Behavior}.\hskip 1em plus 0.5em minus 0.4em\relax Springer, 2006.

\bibitem{paperstuff}
``{Accompanied material and data for this paper},'' \\
  \url{http://claudiomenghi.github.io/RobotPatterns/}, 2018.

\bibitem{ruchkin2018ipl}
I.~Ruchkin, J.~Sunshine, G.~Iraci, B.~Schmerl, and D.~Garlan, ``{IPL}: An
  integration property language for multi-model cyber-physical systems,'' in
  \emph{International Symposium on Formal Methods}.\hskip 1em plus 0.5em minus
  0.4em\relax Springer, 2018.

\bibitem{4209779}
P.~Ulam, Y.~Endo, A.~Wagner, and R.~Arkin, ``Integrated mission specification
  and task allocation for robot teams - design and implementation,'' in
  \emph{International Conference on Robotics and Automation}.\hskip 1em plus
  0.5em minus 0.4em\relax IEEE, 2007.

\bibitem{pnueli1977temporal}
A.~Pnueli, ``The temporal logic of programs,'' in \emph{Foundations of Computer
  Science}.\hskip 1em plus 0.5em minus 0.4em\relax IEEE, 1977.

\bibitem{ben1983temporal}
M.~Ben-Ari, A.~Pnueli, and Z.~Manna, ``The temporal logic of branching time,''
  \emph{Acta informatica}, 1983.

\bibitem{scholarlist}
``{Google Scholar Robotic Venues},'' \url{https://scholar.google.com/citations?
  view\_op=top\_venues\&hl=en\&vq=eng\_robotics}, 2017.

\bibitem{schillinger2016human}
P.~Schillinger, S.~Kohlbrecher, and O.~von Stryk, ``Human-robot collaborative
  high-level control with application to rescue robotics,'' in
  \emph{International Conference on Robotics and Automation (ICRA)}.\hskip 1em
  plus 0.5em minus 0.4em\relax IEEE, 2016.

\bibitem{smith2011optimal}
S.~L. Smith, J.~Tumov{\'a}, C.~Belta, and D.~Rus, ``Optimal path planning for
  surveillance with temporal-logic constraints,'' \emph{The International
  Journal of Robotics Research}, vol.~30, no.~14, pp. 1695--1708, 2011.

\bibitem{Summit}
Robotnik. \url{https://www.robotnik.eu/mobile-robots/summit-xl-hl/}. Accessed:
  2018-06-20.

\bibitem{PA10}
Mitsubishi.
  \url{https://robotik.dfki-bremen.de/en/research/robot-systems/mitsubishi-pa-10-7c.html}.
  Accessed: 2018-06-20.

\bibitem{Tiago}
P.~robotics. \url{http://tiago.pal-robotics.com/}. Accessed: 2018-06-20.

\bibitem{Examples}
Syntech. \url{http://smlab.cs.tau.ac.il/syntech/lego/}. Accessed: 2018-06-20.

\bibitem{spot}
A.~Duret-Lutz, A.~Lewkowicz, A.~Fauchille, T.~Michaud, E.~Renault, and L.~Xu,
  ``Spot 2.0 --- a framework for {LTL} and $\omega$-automata manipulation,'' in
  \emph{Automated Technology for Verification and Analysis}.\hskip 1em plus
  0.5em minus 0.4em\relax Springer, 2016.

\bibitem{tabakov2005experimental}
D.~Tabakov and M.~Y. Vardi, ``{Experimental Evaluation of Classical Automata
  Constructions},'' in \emph{International Conference on Logic for Programming
  Artificial Intelligence and Reasoning}.\hskip 1em plus 0.5em minus
  0.4em\relax Springer, 2005.

\bibitem{de2006antichains}
M.~De~Wulf, L.~Doyen, T.~A. Henzinger, and J.-F. Raskin, ``{Antichains: A New
  Algorithm for Checking Universality of Finite Automata},'' in
  \emph{International Conference on Computer Aided Verification}.\hskip 1em
  plus 0.5em minus 0.4em\relax Springer, 2006.

\bibitem{tabakov2007model}
D.~Tabakov and M.~Y. Vardi, ``{Model Checking Buchi Specifications},'' in
  \emph{International Conference on Language and Automata Theory and
  Applications}, 2007.

\bibitem{saadatpanah2012comparing}
P.~Saadatpanah, M.~Famelis, J.~Gorzny, N.~Robinson, M.~Chechik, and R.~Salay,
  ``{Comparing the Effectiveness of Reasoning Formalisms for Partial Models},''
  in \emph{Workshop on Model-Driven Engineering, Verification and
  Validation}.\hskip 1em plus 0.5em minus 0.4em\relax ACM, 2012.

\bibitem{famelis2012partial}
M.~Famelis, R.~Salay, and M.~Chechik, ``{Partial Models: Towards Modeling and
  Reasoning with Uncertainty},'' in \emph{International Conference on Software
  Engineering (ICSE)}.\hskip 1em plus 0.5em minus 0.4em\relax IEEE, 2012.

\bibitem{FM2016}
C.~Menghi, P.~Spoletini, and C.~Ghezzi, ``Dealing with incompleteness in
  automata-based model checking,'' in \emph{Formal Methods (FM)}.\hskip 1em
  plus 0.5em minus 0.4em\relax Springer, 2016.

\bibitem{fase2018}
C.~Menghi, P.~Spoletini, M.~Chechik, and C.~Ghezzi, ``Supporting
  verification-driven incremental distributed design of components,'' in
  \emph{Fundamental Approaches to Software Engineering}.\hskip 1em plus 0.5em
  minus 0.4em\relax Springer, 2018.

\bibitem{DBLP:conf/icse/MenghiGPT18}
C.~Menghi, S.~Garc{\'{\i}}a, P.~Pelliccione, and J.~Tumova, ``Towards
  multi-robot applications planning under uncertainty,'' in \emph{International
  Conference on Software Engineering (ICSE): Companion Proceeeding}, 2018.

\bibitem{best2016multi}
G.~Best, J.~Faigl, and R.~Fitch, ``Multi-robot path planning for budgeted
  active perception with self-organising maps,'' in \emph{International
  Conference on Intelligent Robots and Systems (IROS)}.\hskip 1em plus 0.5em
  minus 0.4em\relax IEEE, 2016.

\bibitem{takacs2009multi}
B.~Tak{\'a}cs and Y.~Demiris, ``Multi-robot plan adaptation by constrained
  minimal distortion feature mapping,'' in \emph{International Conference on
  Robotics and Automation (ICRA)}.\hskip 1em plus 0.5em minus 0.4em\relax IEEE,
  2009.

\bibitem{stentz1996map}
A.~Stentz, ``Map-based strategies for robot navigation in unknown
  environments,'' in \emph{AAAI spring symposium on planning with incomplete
  information for robot problems}, 1996, pp. 110--116.

\bibitem{OrderSpecificationPatterns}
``{Order Specification Patterns},''
  \url{http://patterns.projects.cs.ksu.edu/docum
  entation/patterns/order.shtml}.

\bibitem{Paun99}
D.~O. Paun and M.~Chechik, ``Events in linear-time properties,'' in
  \emph{International Symposium on Requirements Engineering}.\hskip 1em plus
  0.5em minus 0.4em\relax IEEE, 1999.

\bibitem{Remenska2014}
D.~Remenska, T.~A.~C. Willemse, J.~Templon, K.~Verstoep, and H.~Bal, ``Property
  specification made easy: Harnessing the power of model checking in uml
  designs,'' in \emph{International Federated Conference on Distributed
  Computing Techniques}.\hskip 1em plus 0.5em minus 0.4em\relax Springer, 2014.

\bibitem{Castillos2013}
K.~C. Castillos, F.~Dadeau, J.~Julliand, B.~Kanso, and S.~Taha, \emph{A
  Compositional Automata-Based Semantics for Property Patterns}.\hskip 1em plus
  0.5em minus 0.4em\relax Springer, 2013.

\bibitem{Bitsch2001}
F.~Bitsch, ``Safety patterns - the key to formal specification of safety
  requirements,'' in \emph{International Conference on Computer Safety,
  Reliability and Security}, 2001.

\bibitem{Spanoudakis2007}
G.~Spanoudakis, C.~Kloukinas, and K.~Androutsopoulos, ``Towards security
  monitoring patterns,'' in \emph{Symposium on Applied Computing}.\hskip 1em
  plus 0.5em minus 0.4em\relax ACM, 2007.

\bibitem{Schmidt2006}
D.~C. Schmidt, ``Guest editor's introduction: Model-driven engineering,''
  \emph{Computer}, vol.~39, no.~2, pp. 25--31, 2006.

\bibitem{Ciccozzi4496}
F.~Ciccozzi, D.~D. Ruscio, I.~Malavolta, and P.~Pelliccione, ``Adopting {MDE}
  for specifying and executing civilian missions of mobile multi-robot
  systems,'' \emph{Journal of IEEE Access}, vol.~2, no.~1, 2016.

\bibitem{Adam2014}
S.~Adam, M.~Larsen, K.~Jensen, and U.~P. Schultz, \emph{Towards Rule-Based
  Dynamic Safety Monitoring for Mobile Robots}.\hskip 1em plus 0.5em minus
  0.4em\relax Springer, 2014, pp. 207--218.

\bibitem{aiello2007handbook}
M.~Aiello, I.~Pratt-Hartmann, J.~van Benthem \emph{et~al.}, \emph{Handbook of
  spatial logics}.\hskip 1em plus 0.5em minus 0.4em\relax Springer, 2007,
  vol.~4.

\bibitem{papadimitriou1996topological}
C.~H. Papadimitriou, D.~Suciu, and V.~Vianu, ``Topological queries in spatial
  databases,'' in \emph{Symposium on Principles of database systems}.\hskip 1em
  plus 0.5em minus 0.4em\relax ACM, 1996.

\bibitem{bivand2013spatial}
R.~S. Bivand, E.~Pebesma, and V.~G{\'o}mez-Rubio, \emph{Spatial Data Import and
  Export}.\hskip 1em plus 0.5em minus 0.4em\relax Springer, 2013.

\bibitem{erwtimatiko}
R.~Kontchakov, A.~Kurucz, F.~Wolter, and M.~Zakharyaschev, ``Spatial logic+
  temporal logic=?'' in \emph{Handbook of spatial logics}.\hskip 1em plus 0.5em
  minus 0.4em\relax Springer, 2007, pp. 497--564.

\bibitem{cardelli2002spatial}
L.~Cardelli, P.~Gardner, and G.~Ghelli, ``A spatial logic for querying
  graphs,'' in \emph{Automata, Languages and Programming}.\hskip 1em plus 0.5em
  minus 0.4em\relax Springer, 2002.

\end{thebibliography}

\end{document}